\documentclass[aps,prl,twocolumn,showpacs,preprintnumbers,amsmath,amssymb,refmerge,superscriptaddress]{revtex4}
\usepackage{longtable}
\usepackage{amsmath}
\usepackage{graphicx} % Include figure files
\usepackage{dcolumn}  % Align table columns on decimal point
\usepackage{color}

\renewcommand{\arraystretch}{1.1}

\newcommand{\cm}{\mathrm{cm}}
\newcommand{\mev}{\mathrm{MeV}}

\newcommand{\mevm}{\mathrm{MeV}/c^2}
\newcommand{\gev}{\mathrm{GeV}}
\newcommand{\gevc}{\mathrm{GeV}/c}
\newcommand{\gevm}{\mathrm{GeV}/c^2}
\newcommand{\gevms}{\mathrm{GeV}^2/c^4}
\newcommand{\ee}{e^{+} e^{-}}
\newcommand{\mumu}{\mu^{+}\mu^{-}}

\newcommand{\leplep}{\ell^{+}\ell^{-}}
\newcommand{\jp}{J/\psi}
\newcommand{\ch}{\chi_{c1}}
\newcommand{\z}{Z^+}
\newcommand{\kst}{K^{*}}
\newcommand{\B}{\bar{B^0}}

\newcommand{\km}{K^-}
\newcommand{\pip}{\pi^{+}}
\newcommand{\mbc}{M_{\rm bc}}
\newcommand{\de}{\Delta E}
\newcommand{\fb}{\mathrm{fb}^{-1}}
\newcommand{\br}{\mathcal{B}}

\newcommand{\sx}{s_x}
\newcommand{\sy}{s_y}
\newcommand{\mone}{4051\pm14{^{+20}_{-41}}}
\newcommand{\gone}{82^{+21}_{-17}{^{+47}_{-22}}}
\newcommand{\bone}{(3.0^{+1.5}_{-0.8}{^{+3.7}_{-1.6}})\times10^{-5}}
\newcommand{\mtwo}{4248^{+44}_{-29}{^{+180}_{-\phantom{1}35}}}
\newcommand{\gtwo}{177^{+54}_{-39}{^{+316}_{-\phantom{1}61}}}
\newcommand{\btwo}{(4.0^{+2.3}_{-0.9}{^{+19.7}_{-\phantom{1}0.5}})\times10^{-5}}
\newcommand{\bkmpipch}{(3.83\pm0.10\pm0.39)\times10^{-4}}
\newcommand{\bkstch}{(1.73^{+0.15}_{-0.12}{^{+0.34}_{-0.22}})\times10^{-4}}

\begin{document}

\title{ \quad\\[0.5cm]
Observation of two resonance-like structures in the $\pip\ch$ mass
distribution in exclusive $\B\to\km\pip\ch$ decays }

\begin{abstract}
\noindent
We report the first observation of two resonance-like structures in
the $\pip\ch$ invariant mass distribution near $4.1\,\gevm$ in
exclusive \mbox{$\B\to\km\pip\ch$} decays.
From a Dalitz plot analysis in which the $\pip\ch$ mass structures are
represented by Breit-Wigner resonance amplitudes, we determine masses
and widths of: $M_1 =(\mone)\,\mevm $, $\Gamma_1 =(\gone)\,\mev $,
$M_2 =(\mtwo)\,\mevm $, and $\Gamma_2 =(\gtwo)\,\mev $; and product
branching fractions of
$\br(\B\to\km\z_{1,2})\times\br(\z_{1,2}\to\pip\ch)=\bone$ and
$\btwo$ respectively.
Here the first uncertainty is statistical, the second is systematic.
The significance of each of the $\pip\ch$ structures exceeds
$5\,\sigma$, including the systematic uncertainty from various fit
models.
This analysis is based on $657\times10^6$ $B\bar{B}$ events
collected at the $\Upsilon(4S)$ resonance with the Belle detector at
the KEKB asymmetric-energy $e^+e^-$ collider.
\end{abstract}


\affiliation{Budker Institute of Nuclear Physics, Novosibirsk}
%%%\affiliation{Chiba University, Chiba}
\affiliation{University of Cincinnati, Cincinnati, Ohio 45221}
%%%\affiliation{T. Ko\'{s}ciuszko Cracow University of Technology, Krakow}
\affiliation{Department of Physics, Fu Jen Catholic University, Taipei}
\affiliation{Justus-Liebig-Universit\"at Gie\ss{}en, Gie\ss{}en}
\affiliation{The Graduate University for Advanced Studies, Hayama}
\affiliation{Gyeongsang National University, Chinju}
\affiliation{Hanyang University, Seoul}
\affiliation{University of Hawaii, Honolulu, Hawaii 96822}
\affiliation{High Energy Accelerator Research Organization (KEK), Tsukuba}
%%%\affiliation{Hiroshima Institute of Technology, Hiroshima}
%%%\affiliation{University of Illinois at Urbana-Champaign, Urbana, Illinois 61801}
\affiliation{Institute of High Energy Physics, Chinese Academy of Sciences, Beijing}
\affiliation{Institute of High Energy Physics, Vienna}
\affiliation{Institute of High Energy Physics, Protvino}
\affiliation{Institute for Theoretical and Experimental Physics, Moscow}
\affiliation{J. Stefan Institute, Ljubljana}
\affiliation{Kanagawa University, Yokohama}
\affiliation{Korea University, Seoul}
%%%\affiliation{Kyoto University, Kyoto}
\affiliation{Kyungpook National University, Taegu}
\affiliation{\'Ecole Polytechnique F\'ed\'erale de Lausanne (EPFL), Lausanne}
\affiliation{Faculty of Mathematics and Physics, University of Ljubljana, Ljubljana}
\affiliation{University of Maribor, Maribor}
\affiliation{University of Melbourne, School of Physics, Victoria 3010}
\affiliation{Nagoya University, Nagoya}
\affiliation{Nara Women's University, Nara}
\affiliation{National Central University, Chung-li}
\affiliation{National United University, Miao Li}
\affiliation{Department of Physics, National Taiwan University, Taipei}
\affiliation{H. Niewodniczanski Institute of Nuclear Physics, Krakow}
\affiliation{Nippon Dental University, Niigata}
\affiliation{Niigata University, Niigata}
\affiliation{University of Nova Gorica, Nova Gorica}
\affiliation{Osaka City University, Osaka}
%%%\affiliation{Osaka University, Osaka}
\affiliation{Panjab University, Chandigarh}
%%%\affiliation{Peking University, Beijing}
%%%\affiliation{Princeton University, Princeton, New Jersey 08544}
%%%\affiliation{RIKEN BNL Research Center, Upton, New York 11973}
\affiliation{Saga University, Saga}
\affiliation{University of Science and Technology of China, Hefei}
\affiliation{Seoul National University, Seoul}
%%%\affiliation{Shinshu University, Nagano}
\affiliation{Sungkyunkwan University, Suwon}
\affiliation{University of Sydney, Sydney, New South Wales}
%%%\affiliation{Tata Institute of Fundamental Research, Mumbai}
\affiliation{Toho University, Funabashi}
\affiliation{Tohoku Gakuin University, Tagajo}
%%%\affiliation{Tohoku University, Sendai}
\affiliation{Department of Physics, University of Tokyo, Tokyo}
\affiliation{Tokyo Institute of Technology, Tokyo}
\affiliation{Tokyo Metropolitan University, Tokyo}
\affiliation{Tokyo University of Agriculture and Technology, Tokyo}
%%%\affiliation{Toyama National College of Maritime Technology, Toyama}
\affiliation{Virginia Polytechnic Institute and State University, Blacksburg, Virginia 24061}
\affiliation{Yonsei University, Seoul}
   \author{R.~Mizuk}\affiliation{Institute for Theoretical and Experimental Physics, Moscow} % ITEP
   \author{R.~Chistov}\affiliation{Institute for Theoretical and Experimental Physics, Moscow} % ITEP
   \author{I.~Adachi}\affiliation{High Energy Accelerator Research Organization (KEK), Tsukuba} % KEK
   \author{H.~Aihara}\affiliation{Department of Physics, University of Tokyo, Tokyo} % Tokyo
% \author{D.~Anipko}\affiliation{Budker Institute of Nuclear Physics, Novosibirsk} % BINP
   \author{K.~Arinstein}\affiliation{Budker Institute of Nuclear Physics, Novosibirsk} % BINP
% \author{T.~Aso}\affiliation{Toyama National College of Maritime Technology, Toyama} % Toyama
   \author{V.~Aulchenko}\affiliation{Budker Institute of Nuclear Physics, Novosibirsk} % BINP
   \author{T.~Aushev}\affiliation{\'Ecole Polytechnique F\'ed\'erale de Lausanne (EPFL), Lausanne}\affiliation{Institute for Theoretical and Experimental Physics, Moscow} % ITEP
% \author{T.~Aziz}\affiliation{Tata Institute of Fundamental Research, Mumbai} % Tata
% \author{S.~Bahinipati}\affiliation{University of Cincinnati, Cincinnati, Ohio 45221} % Cincinnati
   \author{A.~M.~Bakich}\affiliation{University of Sydney, Sydney, New South Wales} % Sydney
   \author{V.~Balagura}\affiliation{Institute for Theoretical and Experimental Physics, Moscow} % ITEP
% \author{Y.~Ban}\affiliation{Peking University, Beijing} % Peking
   \author{E.~Barberio}\affiliation{University of Melbourne, School of Physics, Victoria 3010} % Melbourne
% \author{M.~Barbero}\affiliation{University of Hawaii, Honolulu, Hawaii 96822} % Hawaii
% ...
% BINP
\collaboration{The Belle Collaboration}

\maketitle

\tighten

{\renewcommand{\thefootnote}{\fnsymbol{footnote}}}
\setcounter{footnote}{0}

\section{Introduction}

Recently the Belle Collaboration reported the observation of a
relatively narrow resonance-like structure in the $\pip\psi(2S)$ mass
spectrum produced in $\B\to\km\pip\psi(2S)$ decays, calling this
structure the $Z(4430)^+$\cite{Z4430}. If the $Z(4430)^+$ is
interpreted as a meson state, then its minimal quark content must be
the exotic combination $|c\bar{c}u\bar{d}\rangle$.  The $Z(4430)^+$
observation motivated studies of other $\B\to\km\pip(c\bar{c})$
decays.

In this paper we present a study of the decay $\B\to\km\pip\ch$,
including the first observation of a doubly peaked structure in the
$\pip\ch$ invariant mass distribution near $4.1\,\gevm$. If the two
peaks are meson states, their minimal quark content must be the same
as that of the $Z(4430)^+$.
The analysis is performed using data collected with the Belle detector
at the KEKB asymmetric-energy $e^+e^-$ collider~\cite{kekb}.  The data
sample consists of $605\,\fb$ accumulated at the $\Upsilon(4S)$
resonance, which corresponds to $657\times10^6$ $B\bar{B}$ pairs.

\section{Belle Detector}

The Belle detector is a large-solid-angle magnetic spectrometer that
consists of a silicon vertex detector (SVD), a 50-layer central drift
chamber (CDC), an array of aerogel threshold Cherenkov counters (ACC),
a barrel-like arrangement of time-of-flight scintillation counters
(TOF), and an electromagnetic calorimeter (ECL) comprising CsI(Tl)
crystals located inside a superconducting solenoid coil that provides
a 1.5$\,$T magnetic field.  An iron flux-return located outside of the
coil is instrumented to detect $K^0_L$ mesons and to identify muons
(KLM). The detector is described in detail
elsewhere~\cite{BELLE_DETECTOR}.  Two different inner detector
configurations were used, a $2.0\,\cm$ radius beam-pipe and a 3-layer
silicon vertex detector for the first $155\,\fb$, and a $1.5\,\cm$
radius beam-pipe with a 4-layer vertex detector for the remaining
$450\,\fb$~\cite{svd2}.

We use a GEANT-based Monte Carlo (MC) simulation~\cite{geant} to model
the response of the detector, identify potential backgrounds and
determine the acceptance. The MC simulation includes run-dependent
detector performance variations and background conditions.  Signal MC
events are generated in proportion to the relative luminosities of the
different running periods.

\section{Event Selection}

We select events of the type $\B\to\km\pip\ch$, where the $\ch$ meson
is reconstructed via its decay to $\jp\gamma$, with a subsequent $\jp$
decay to $\leplep$ ($\leplep = \ee$ or $\mumu$). The inclusion of
charge-conjugate modes is implied throughout this paper.

All tracks are required to originate from the beam-beam interaction
region: $dr<0.2\,\cm$ and $dz<2\,\cm$, where $dr$ is the distance of
closest approach to the beam-beam interaction point in the plane
perpendicular to the beam axis and $dz$ is the corresponding distance
along the beam direction.
Charged pions and kaons are identified using a likelihood ratio method
that combines information from the TOF system and ACC counters with
energy loss ($dE/dx$) measurements from the CDC.  The identification
requirements for kaons have an efficiency of 90\% and a pion
misidentification probability of 10\%.
Muons are identified by their range and transverse scattering in the
KLM. Electrons are identified by the presence of a matching ECL
cluster with transverse energy profile consistent with an
electromagnetic shower.
In addition, charged pions and kaons that are also positively
identified as electrons are rejected.

Photons are identified as energy clusters in the ECL that have no
associated charged tracks detected in the CDC, and a shower shape that
is consistent with that of a photon.

For $\jp\to\ee$ candidates, photons that have laboratory frame
energies greater than $30\,\mev$ and are within 50$\,$mrad of the
direction of the $e^+$ or $e^-$ tracks are included in the invariant
mass calculation; we require $|M(\ee)-m_{\jp}|<50\,\mevm$.
For $\jp\to\mumu$ candidates we require
$|M(\mumu)-m_{\jp}|<30\,\mevm$.
To enhance the precision of the $\jp$ energy and momentum
determination, we perform a mass constrained fit to the $\jp$
candidates.

For $\ch\to\jp\gamma$ candidates, we use photons with laboratory frame
energies greater than $50\,\mev$ and require
$|M(\jp\gamma)-m_{\ch}|<30\,\mevm$.
To improve the accuracy of the $\ch$ energy and momentum
determination, we perform a mass constrained fit to the $\ch$
candidates.

Candidate $\B\to\km\pip\ch$ decays are identified by their
center-of-mass (c.m.)\ energy difference, $\Delta
E=\Sigma_iE_i-E_{\mathrm{beam}}$, and their beam-energy constrained
mass,
$M_{\mathrm{bc}}=\sqrt{E^2_{\mathrm{beam}}-(\Sigma_i\vec{p}_i)^2}$,
where $E_{\mathrm{beam}}=\sqrt{s}/2$ is the beam energy in the
c.m.\ and $\vec{p}_i$ and $E_i$ are the three-momenta and energies of
the $B$ candidate's decay products.  We accept $B$ candidates with
$5275\,\mevm<\mbc<5287\,\mevm$ and $|\de|<12\,\mev$. The $\de$
sidebands are defined as $24\,\mev<|\de|<96\,\mev$.
To have well defined Dalitz plot boundaries for both signal and
sideband events, we perform a mass constrained fit to the $\B$
candidates from both regions (to the nominal $\B$ mass in all cases).

\section{Analysis of $\B\to\km\pip\ch$ Decays }

\begin{figure}[htbp]
\includegraphics[width=6cm]{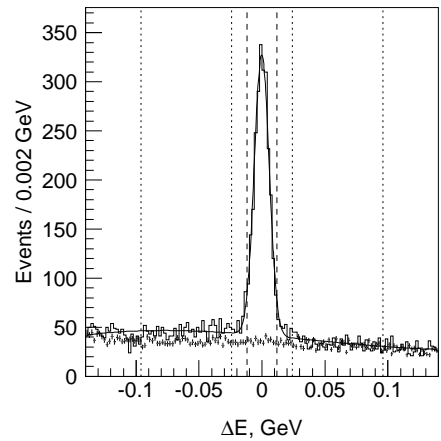}
\caption{ The $\de$ distribution for the selected $\B$ meson
  candidates (histogram) and for the $\ch$ sidebands (points with
  error bars). The vertical lines indicate the $\de$ signal and
  sideband regions.}
\label{de_fit}
\end{figure}

\begin{figure}[htbp]
\includegraphics[width=6cm]{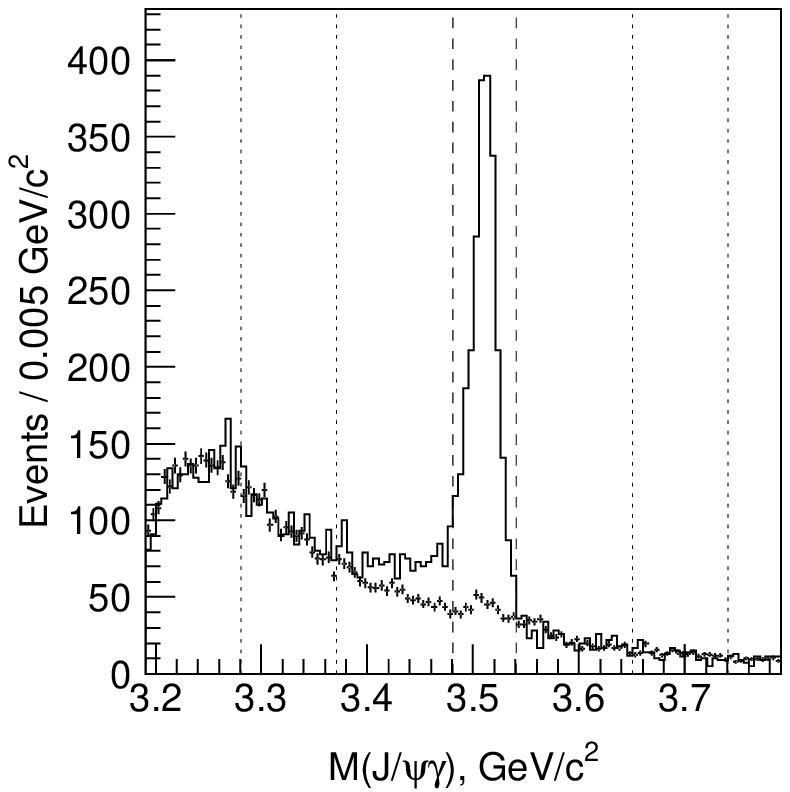}
\caption{ The $M(\jp\gamma)$ distribution for the selected $\B$ meson
  candidates (histogram) and for the $\de$ sidebands (points with
  error bars). The vertical lines indicate the $\ch$ signal and
  sideband regions. }
\label{mhic}
\end{figure}

The $\de$ distribution for selected $\B\to\km\pip\ch$ candidates is
shown in Fig.~\ref{de_fit}.
The contribution of the $\ch$ sideband regions defined as
$140\,\mevm<|M(\jp\gamma)-m_{\ch}|<230\,\mevm$ is also shown. The
$\ch$ sidebands account for almost all the background, which indicates
that the background is primarily due to combinatorial photons; the
contamination from events with misidentified particles is found to be
negligibly small.
The $M(\jp\gamma)$ distributions before the $\ch$ mass constrained fit
for the $\de$ signal and sideband regions are shown in
Fig.~\ref{mhic}.
There is a small $\ch$ signal in the $\de$ sidebands due to inclusive
$\ch$ production in $B$ decays.
The $\jp$ signals in the $M(\mu^+\mu^-)$ and $M(e^+e^-)$ distributions
are almost background-free.

A signal yield of $2126 \pm 56 \pm 42$ $\B\to\km\pip\ch$ events is
determined from a fit to the $\de$ distribution using a Gaussian
function to represent the signal plus a second-order polynomial to
represent the background.  The fitted $\de$ resolution,
$\sigma=(5.93\pm0.15\pm0.13)\,\mevm$, is consistent with the MC
expectation of $\sigma=(5.62\pm0.03\pm0.09)\,\mevm$.  Here and
elsewhere in this report the first uncertainty is statistical, the
second is systematic. The systematic uncertainties for the signal
yield and the $\de$ width are estimated by varying the $\de$ interval
covered by the fit.

To determine the detection efficiency, we simulate $B^0\B$ events
where $\B\to\km\pip\ch$ with a uniform phase-space distribution and
the accompanying $B^0$ decays generically.  These MC events are then
weighted according to the results of the fit to the Dalitz plot that
is described below. In this way, the reconstruction efficiency is
found to be $(20.0\pm1.4)\%$, where the following sources are included
in the uncertainty: the dependence on the Dalitz plot model (0.2\%);
data and MC differences for track and $\gamma$ reconstruction
($1\%\times4$ for four tracks and 1.5\% for $\gamma$), and particle
identification (4\% for the $\km\pip$ pair and 4.2\% for $\leplep$);
uncertainties in the angular distributions for $\ch\to\jp\gamma$ and
$\jp\to\leplep$ decays (0.2\%); and MC statistics (0.6\%). The
uncertainties from different sources are added in quadrature.
The efficiency is corrected for the difference in lepton
identification performance in data compared to MC, ($-4.5\pm 4.2)\%$,
as estimated from $\jp\to\leplep$ and $e^+e^-\to e^+e^-\leplep$
control samples.

Using $(656.7\pm8.9)\times10^6$ as the number of $B\bar{B}$ pairs and
Particle Data Group (PDG) 2006 values for the branching fractions
$\br(\ch\to\jp\gamma)=0.356\pm0.019$ and
$\br(\jp\to\leplep)=0.1187\pm0.0012$~\cite{PDG}, we determine
\[
\br(\B\to\km\pip\ch)=\bkmpipch.
\]
The systematic uncertainty includes contributions from the uncertainty
in the efficiency (7.2\%), the systematic uncertainty in the signal
yield (2.0\%), the uncertainty due to the variation in the selection
requirements (3.9\%), the uncertainty in the $\de$ signal shape
(1.0\%, considering a sum of two Gaussian functions instead of a
single one) and the uncertainties in the $\ch$ and $\jp$ decay
branching fractions (5.3\% and 1.0\%, respectively).

\begin{figure}[htbp]
\includegraphics[width=8cm]{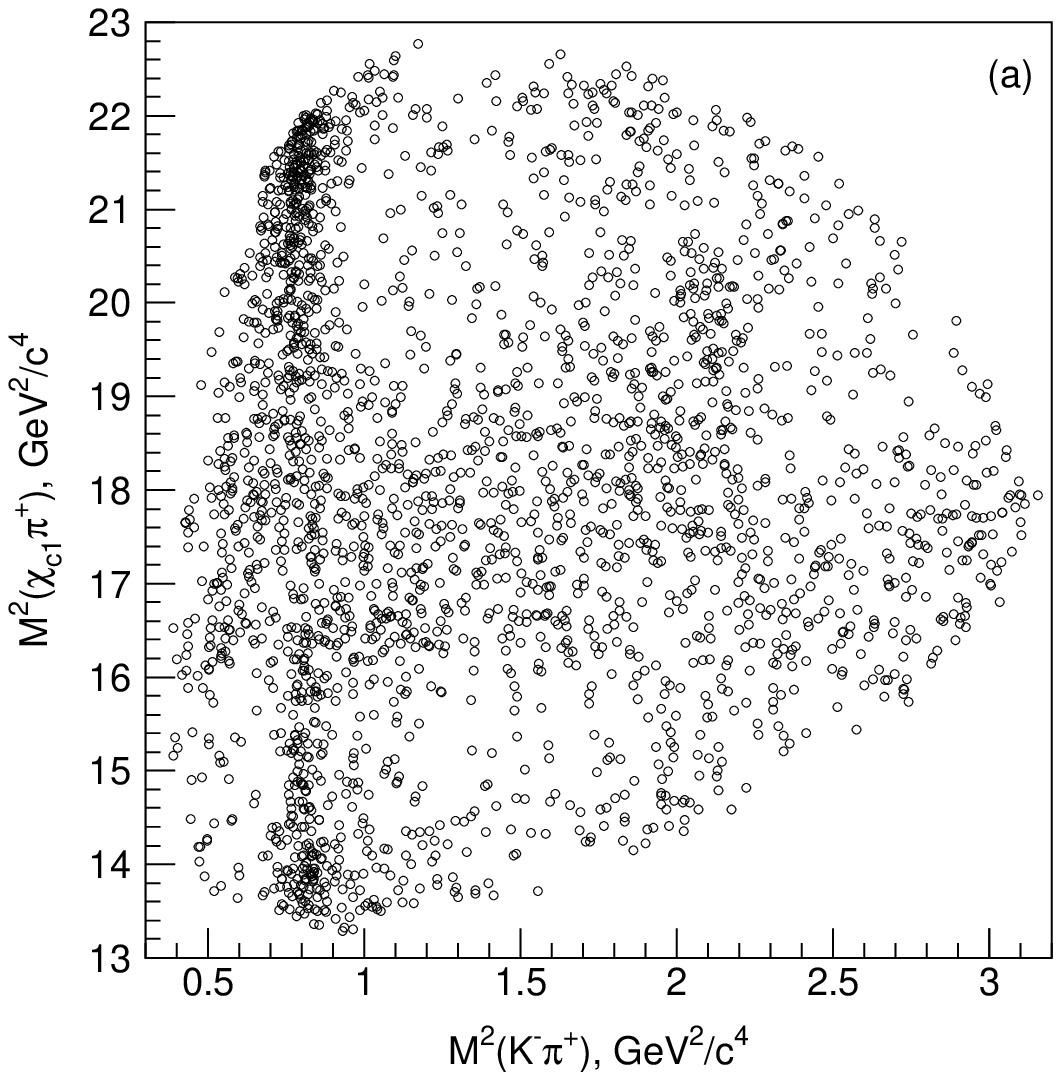}
\includegraphics[width=8cm]{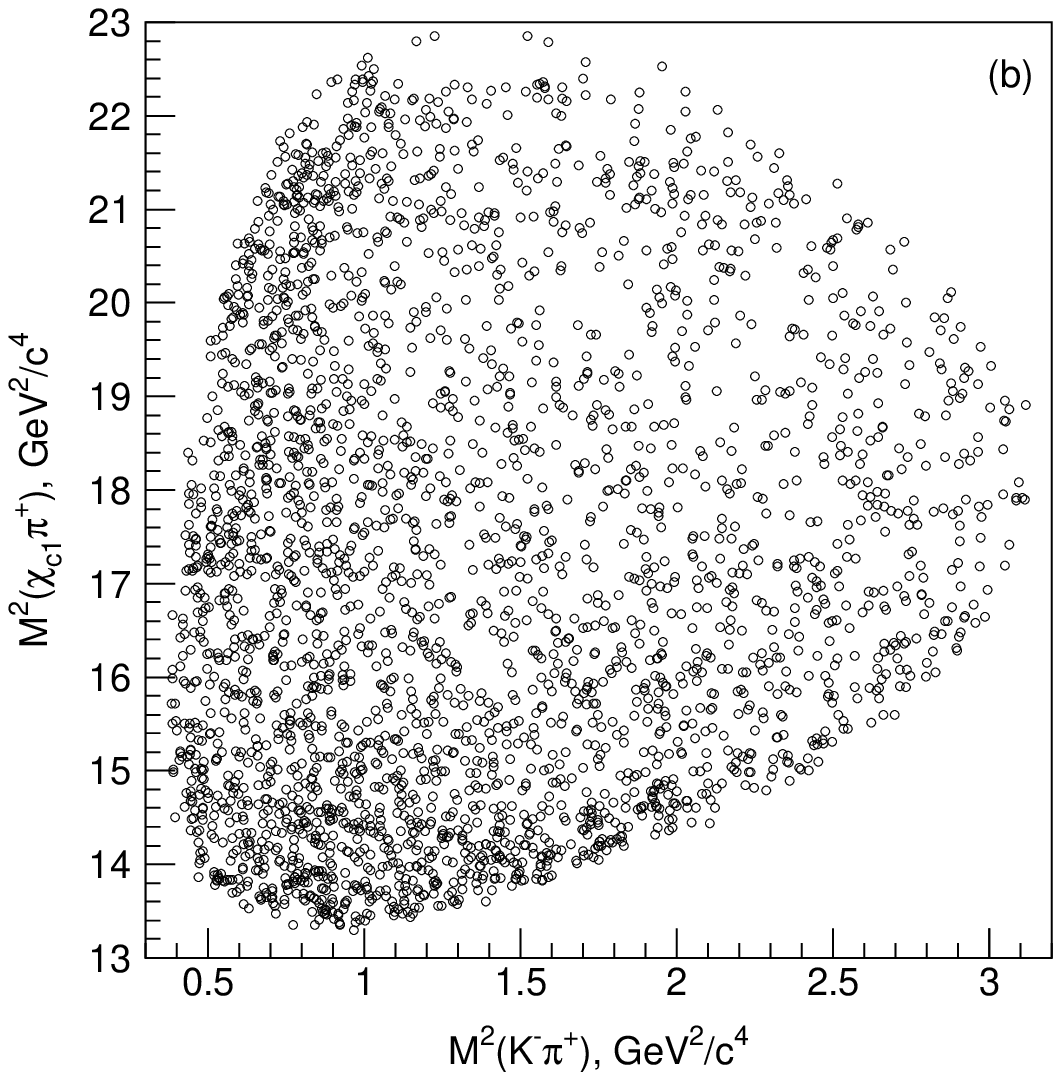}
\caption{ The $\B\to\km\pip\ch$ decay Dalitz plot for the $\de$ signal
 (a) and sideband (b) regions. }
\label{dalitz}
\end{figure}

The $\B\to\km\pip\ch$ decay Dalitz plot ($M^2(\pip\ch)$ {\it versus}
$M^2(\km\pip)$) for the $\de$ signal region is shown in 
Fig.~\ref{dalitz}~(a).
The Dalitz plot distribution exhibits some distinct features:
a vertical band at $M^2(\km\pip)\simeq0.8\,\gevms$ that corresponds to
$\B\to\kst(892)\ch$ decays;
a clustering of events at $M^2(\km\pip)\simeq2\,\gevms$ that
corresponds primarily to $\B\to\kst(1430)\ch$ decays;
a distinct horizontal band at $M^2(\pip\ch)\simeq17\,\gevms$
corresponding to a structure in the $\pip\ch$ channel, denoted by
$\z$. This latter feature is the subject of this report.

\begin{figure}[htbp]
\includegraphics[width=8cm]{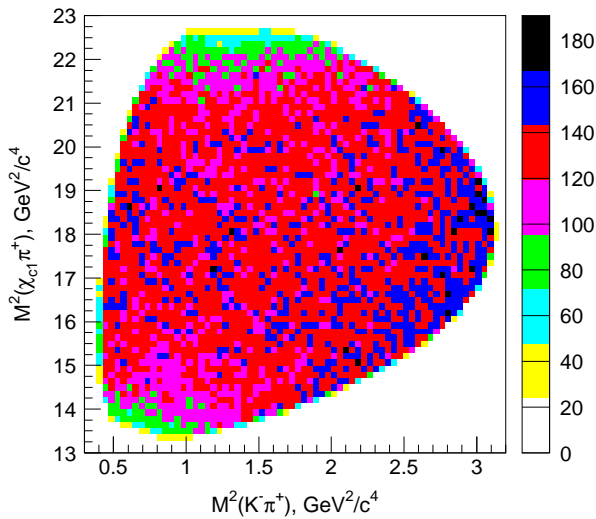}
\caption{ Dalitz plot for reconstructed $\B\to\km\pip\ch$ MC events,
  generated according to the phase space model. }
\label{eff}
\end{figure}

In contrast, the Dalitz plot for the $\de$ sidebands, shown in
Fig.~\ref{dalitz}~(b), is relatively smooth and featureless.  The
Dalitz plot for the phase-space MC candidate events, shown in
Fig.~\ref{eff}, also exhibits a smooth and featureless behaviour.
There is a decrease in efficiency in the top (bottom) region where the
$\km$ ($\pip$) is slow and has a low detection efficiency.

\section{Formalism of Dalitz Analysis}

The decay $\B\to\km\pip\ch$ with the $\ch$ reconstructed in the
$\jp\gamma$ decay mode and the $\jp$ reconstructed in the $\leplep$
decay mode is described by six variables (assuming the widths of the
$\ch$ and $\jp$ to be negligible). We take these to be $M(\pip\ch)$,
$M(\km\pip)$, the $\ch$ and $\jp$ helicity angles ($\theta_{\ch}$ and
$\theta_{\jp}$), and the angle between the $\ch$ ($\jp$) production
and decay planes $\phi_{\ch}$ ($\phi_{\jp}$).  Here we analyze the
$\B\to\km\pip\ch$ decay process after integrating over the angular
variables $\theta_{\ch}$, $\theta_{\jp}$, $\phi_{\ch}$ and
$\phi_{\jp}$. We find that the reconstruction efficiency is almost
uniform over the full $\phi_{\ch}$ and $\phi_{\jp}$ angular ranges;
therefore, after integrating over these angles the interference terms
between different $\ch$ helicity states, which contain factors of
$\sin\phi_{\ch}$, $\cos\phi_{\ch}$, $\sin2\phi_{\ch}$ or
$\cos2\phi_{\ch}$, are negligibly small.  We subsequently verify that
the $\theta_{\ch}$ and $\theta_{\jp}$ distributions agree with these
expectations.

We perform a binned likelihood fit to the Dalitz plot distribution,
where the bin size is chosen by decreasing its area until the fit
results are unaffected by further changes.  The selected number of bins
is $400\times400$.  We consider only those bins that are fully
contained within the Dalitz plot boundaries; this corresponds to
99.3\% of the total Dalitz plot area.

In 1.9\% of events from the $\de$ signal region we find more than one
$\B$ candidate. Multiple candidates are uniformly distributed over the
entire Dalitz plot area. No best candidate selection is applied.

We use a fitting function of the form
\begin{equation}
F(\sx,\sy)=S(\sx,\sy)\times\epsilon(\sx,\sy) + B(\sx,\sy),
\end{equation}
where $\sx\equiv M^2(\km\pip)$, $\sy\equiv M^2(\pip\ch)$, $S$ and $B$
are the signal and background event density functions, and $\epsilon$
is the detection efficiency.
The background $B(\sx,\sy)$ is determined from the $\de$
sidebands. Its normalization is allowed to float in the fit within its
corresponding uncertainty.
The bin-by-bin efficiency $\epsilon(\sx,\sy)$ is determined from the
MC simulation.
Both sidebands and efficiency distributions are smoothed.

The amplitude for the three-body decay $\B\to\km\pip\ch$ is
represented as the sum of Breit-Wigner contributions for different
intermediate two-body states. This type of description, which is
widely used in high energy physics for Dalitz plot
analyses~\cite{cleo}, cannot be exact since it is neither unitary nor
analytic and does not take into account a complete description of
final state interactions.  Nevertheless, the sum of Breit-Wigner terms
reflects the main features of the amplitude's behaviour and provides a
way to find and distinguish the contributions of the two-body
intermediate states, their mutual interference, and their effective
resonance parameters.

Our default fit model includes all known $\km\pip$ resonances below
$1900\,\mevm$ ($\kappa$, $\kst(892)$, $\kst(1410)$, $\kst_0(1430)$,
$\kst_2(1430)$, $\kst(1680)$, $\kst_3(1780)$) and a single exotic
$\ch\pip$ resonance.
The amplitude for $\B\to\km\pip\ch$ via a two-body intermediate
resonance $R$ ($R$ denotes either a $\km\pip$ or $\pip\ch$ resonance)
and the $\ch$ meson in helicity state $\lambda$ is given by
\begin{equation} 
\begin{split}
&A^R_{\lambda}(\sx,\sy)=\\
&F^{(L_B)}_B\cdot\frac{1}{M^2_R-s_R-iM_R\Gamma(s_R)}\cdot F^{(L_R)}_R\cdot\\
&T_{\lambda}\cdot\left(\frac{p_B}{m_B}\right)^{L_B}\cdot
\left(\frac{p_R}{\sqrt{s_R}}\right)^{L_R}.
\end{split}
\end{equation}
Here $F^{(L_B)}_B$ and $F^{(L_R)}_R$ are the $\B$ meson and $R$
resonance decay form factors (the superscript denotes the orbital
angular momentum of the decay); $M_R$ is the resonance mass, $s_R$ is
the four-momentum-squared and $\Gamma(s_R)$ is the energy-dependent
width of the $R$ resonance; $T_{\lambda}$ is the angle-dependent term;
$(\frac{p_B}{m_B})^{L_B}\cdot(\frac{p_R}{\sqrt{s_R}})^{L_R}$ is a
factor related to the momentum dependence of the wave function, $p_B$
($p_R$) is the $\B$ meson ($R$ resonance) daughter's momentum in the
$B$ ($R$) rest frame; and $m_B$ is the $\B$ meson mass.

We use the Blatt-Weisskopf form factors given in
Ref.~\cite{blatt-weisskopf}:
\begin{equation}
\begin{split}
F^{(0)}= & 1 ,\\
F^{(1)}= & \frac{\sqrt{1+z_0}}{\sqrt{1+z}} ,\\
F^{(2)}= & \frac{\sqrt{z_0^2+3z_0+9}}{\sqrt{z^2+3z+9}} ,\\
F^{(3)}= & \frac{\sqrt{z_0^3+6z_0^2+45z_0+225}}{\sqrt{z^3+6z^2+45z+225}}.
\end{split}
\end{equation}
Here $z=r^2p_R^2$ where $r$ is the hadron scale, taken to be
$r=1.6\,(\gevc)^{-1}$, and $z_0=r^2p_{R0}^2$ where $p_{R0}$ is the $R$
resonance daughter's momentum calculated for the pole mass of the $R$
resonance.
For $\kst$ resonances with non-zero spin, the $B$ decay orbital
angular momentum $L_B$ can take several values ($S$, $P$ \& $D$-waves
for $J=1$; $P$, $D$ \& $F$-waves for $J=2$; and $D$, $F$ \& $G$-waves
for $J=3$).  We take the lowest $L_B$ as the default value and
consider the other possibilities in the systematic uncertainty.  The
energy-dependent width is parameterized as
\begin{equation}
\Gamma(s_R)=\Gamma_0\cdot(p_R/p_{R0})^{2L_R+1}\cdot(m_R/\sqrt{s_R})\cdot F_R^2.
\end{equation}

The angular function $T_{\lambda}$ is obtained using the helicity
formalism.
For the $\B\to\kst(\to\km\pip)\ch$ decay
\begin{equation}
T_{\lambda}=d^{J}_{\lambda\,0}(\theta_{\kst}),
\end{equation}
where $J$ is the spin of the $\kst$ resonance; $\theta_{\kst}$ is the
helicity angle of the $\kst$ decay.
For the \mbox{$\B\to\km\z(\to\pip\ch)$} decay
\begin{equation}
T_{\lambda}=d^{J}_{0\,\lambda}(\theta_{\z}),
\end{equation}
where $J$ is the spin of the $\z$ resonance and $\theta_{\z}$ is the
helicity angle of the $\z$ decay.

\begin{figure}[htbp]
\includegraphics[width=8cm]{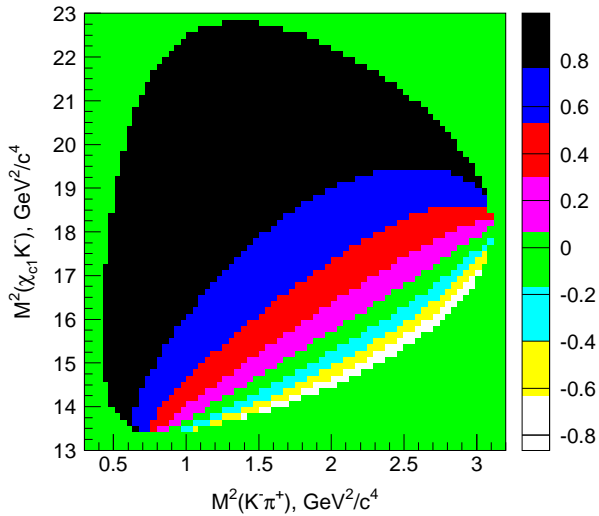}
\caption{ $\cos\theta$ as a function of the Dalitz plot
  variables, where $\theta$ is the angle between the
helicity axes for $K\pi$ and $\pi\ch$ intermediate resonances. }
\label{theta}
\end{figure}

In the decays $\B\to\kst(\to\km\pip)\ch$ and $\B\to\km\z(\to\pip\ch)$
the parent particles of the $\ch$ are different and, therefore, the
relevant $\ch$ helicity is defined relative to different axes:
for $\B\to\kst(\to\km\pip)\ch$ the axis is parallel to the $\km\pip$
momentum in the $\ch$ rest frame; for $\B\to\km\z(\to\pip\ch)$ the
axis is parallel to the $\pip$ momentum in the $\ch$ rest frame. The
angle $\theta$ between the two axes depends upon the event's location
in the Dalitz plot as indicated in Fig.~\ref{theta}.
As a result, the state $|\lambda\rangle_{\z}$ with $\ch$ helicity
$\lambda$ produced in the decay $\B\to\km\z(\to\pip\ch)$ is not equal
to the state $|\lambda\rangle_{\kst}$ with the same $\ch$ helicity
$\lambda$ produced in the decay $\B\to\kst(\to\km\pip)\ch$. The two
states are related by the Wigner $d$-functions via
\begin{equation}
|\lambda\rangle_{\kst} =
\sum_{\lambda'=-1,0,1}{d^1_{\lambda'\lambda}(\theta)}\,|\lambda'\rangle_{\z};
\end{equation}
the same relation holds for the amplitudes.

The resulting expression for the signal event density function is
\begin{equation}
\begin{split}
&S(\sx,\sy)=\\
&\quad\sum_{\lambda=-1,0,1}\Big|\sum_{\kst}\,
a^{\kst}_{\lambda}e^{i\phi^{\kst}_{\lambda}}\,A^{\kst}_{\lambda}(\sx,\sy)+\\
&\qquad\qquad\sum_{\lambda'=-1,0,1}{d^1_{\lambda'\lambda}(\theta)}\,
a^{\z}_{\lambda'}e^{i\phi^{\z}_{\lambda'}}\,A^{\z}_{\lambda'}(\sx,\sy)\Big|^2,
\end{split}
\end{equation}
where $a^R_{\lambda}$ and $\phi^R_{\lambda}$ are the normalizations
and phases of the amplitudes for the intermediate resonance $R$ and
$\ch$ helicity $\lambda$. The phase $\phi^{\kst(892)}_0$ is fixed to
zero.
The detector resolution ($\sigma\sim2\,\mevm$) is small compared to
the width of any of the resonances that are considered and is ignored.

The masses and widths of the $\kst$ resonances are fixed to their PDG
values, except for the $\kappa$, for which the mass and width are
allowed to vary within their experimental
uncertainties~\cite{PDG}. The mass and width of the $\z$ is allowed to
vary without any restrictions.

\section{Results}

\begin{figure}[htbp]
\includegraphics[width=8cm]{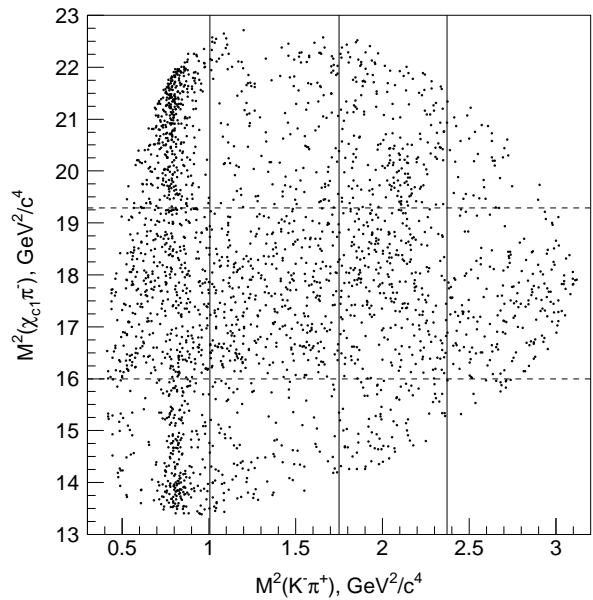}
\caption{ The $\B\to\km\pip\ch$ decay Dalitz plot. The solid (dashed)
  lines delimit the four vertical (three horizontal) slices that are
  used to present the fit results in subsequent figures. The
  coordinates of the vertical (horizontal) lines are
  $M^2(\km\pip)=1.00\,\gevms$, $1.75\,\gevms$ and $2.37\,\gevms$
  ($M^2(\pip\ch)=16.0\,\gevms$ and $19.3\,\gevms$).
}
\label{lines}
\end{figure}

\begin{figure}[htbp]
\includegraphics[width=8cm]{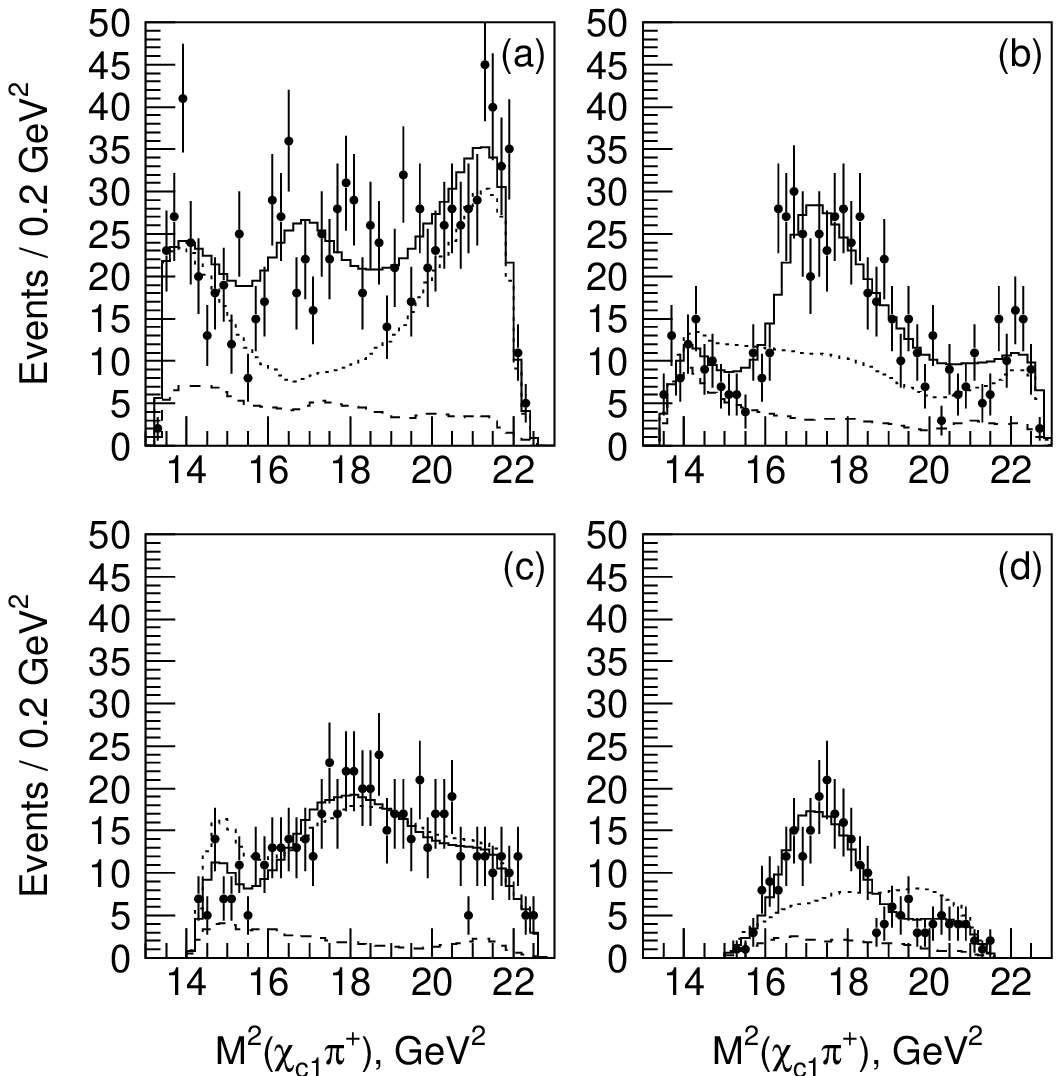}
\includegraphics[width=8cm]{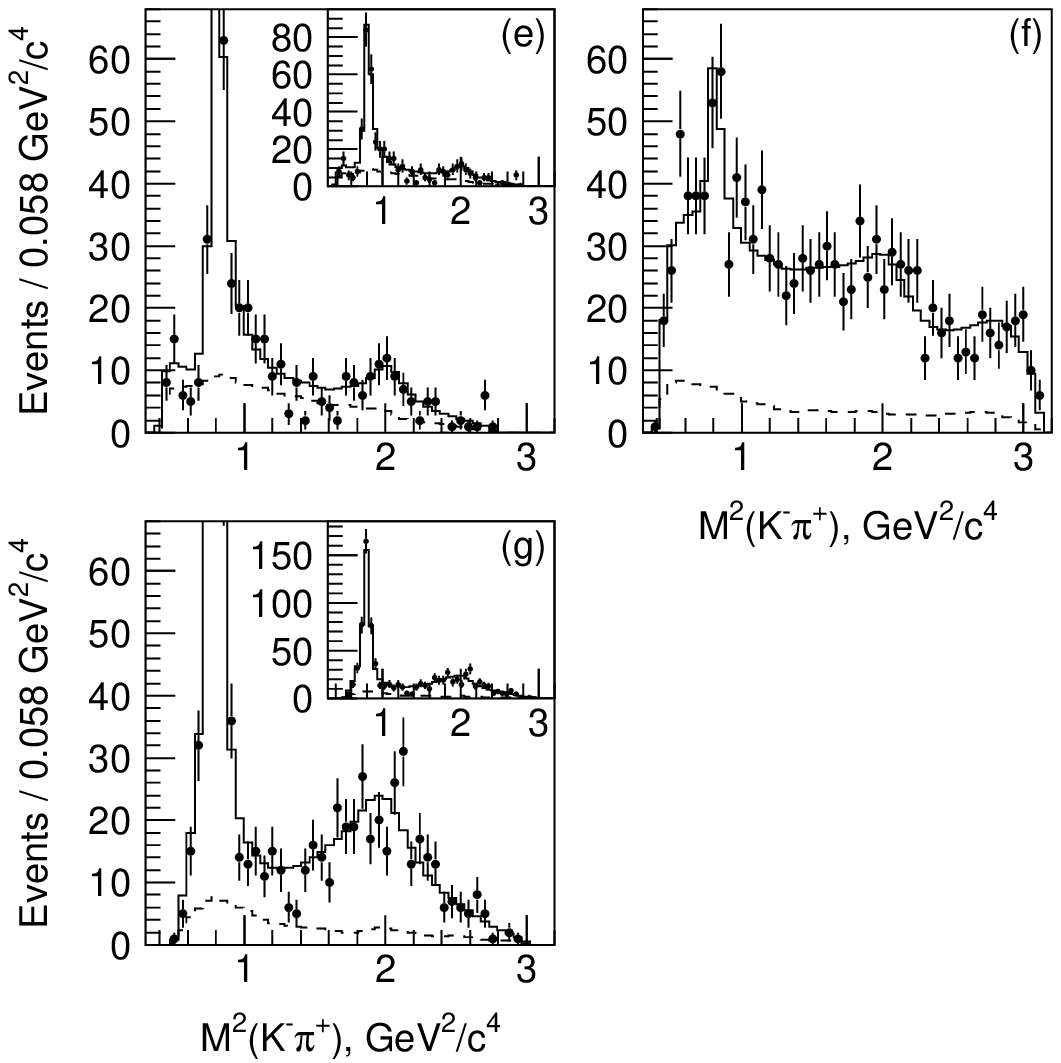}
\caption{ Dalitz plot fit with the default model, including one
  $\z$. Projections for the slices defined in Fig.~\ref{lines} are
  shown: (a)-(d) correspond to vertical slices from left to right,
  (e)-(g) correspond to horizontal slices from bottom to top; in (e)
  and (g), plots including the full vertical scale are shown inset.
  The dots with error bars represent data, the solid histograms are
  the fit results, the dashed histograms represent the background, and
  the dotted histograms in (a)-(d) represent the sum of all fit
  components except the $\z$. The fitting model includes all known
  $\kst$ resonances and one $\z$ term. }
\label{allk_s}
\end{figure}

To display the results of the fit, we divide the Dalitz plot into four
vertical and three horizontal slices as shown in
Fig.~\ref{lines}.
Projections of the fit results for the seven slices are shown in
Fig.~\ref{allk_s}, where the influence of the structure in the
$\pip\ch$ channel is most clearly seen in the second vertical slice.
The mass and width of the $\z$ found from the fit are
$M=(4150^{+31}_{-16})\,\mevm$ and $\Gamma=(352^{+99}_{-43})\,\mev$;
the fit fraction of $\z$ events, defined as the integral of the $\z$
contribution over the Dalitz plot divided by the integral of the
signal function, $\frac{\int |A_z|^2 d\sx d\sy}{\int S \; d\sx d\sy}$,
is $(33.1^{+8.7}_{-5.8})\%$.
All quoted uncertainties are statistical.

The fit fraction is not determined directly from the fit and its
statistical uncertainty is difficult to estimate based on the
statistical uncertainties of fit parameters. 
In this paper the statistical uncertainties of fit fractions are
determined using 1000 toy Monte Carlo samples. Each sample is
generated according to the probability distribution determined from
the fit to experimental data, and contains the same number of events
as the data. We generate 1000 such samples, fit them, and determine
the fit fractions. We fit the distribution of the obtained fit
fractions to an asymmetric Gaussian function and consider the sigmas
of the Gaussian function as the statistical uncertainty in the fit
fraction.

\begin{figure}[htbp]
\includegraphics[width=8cm]{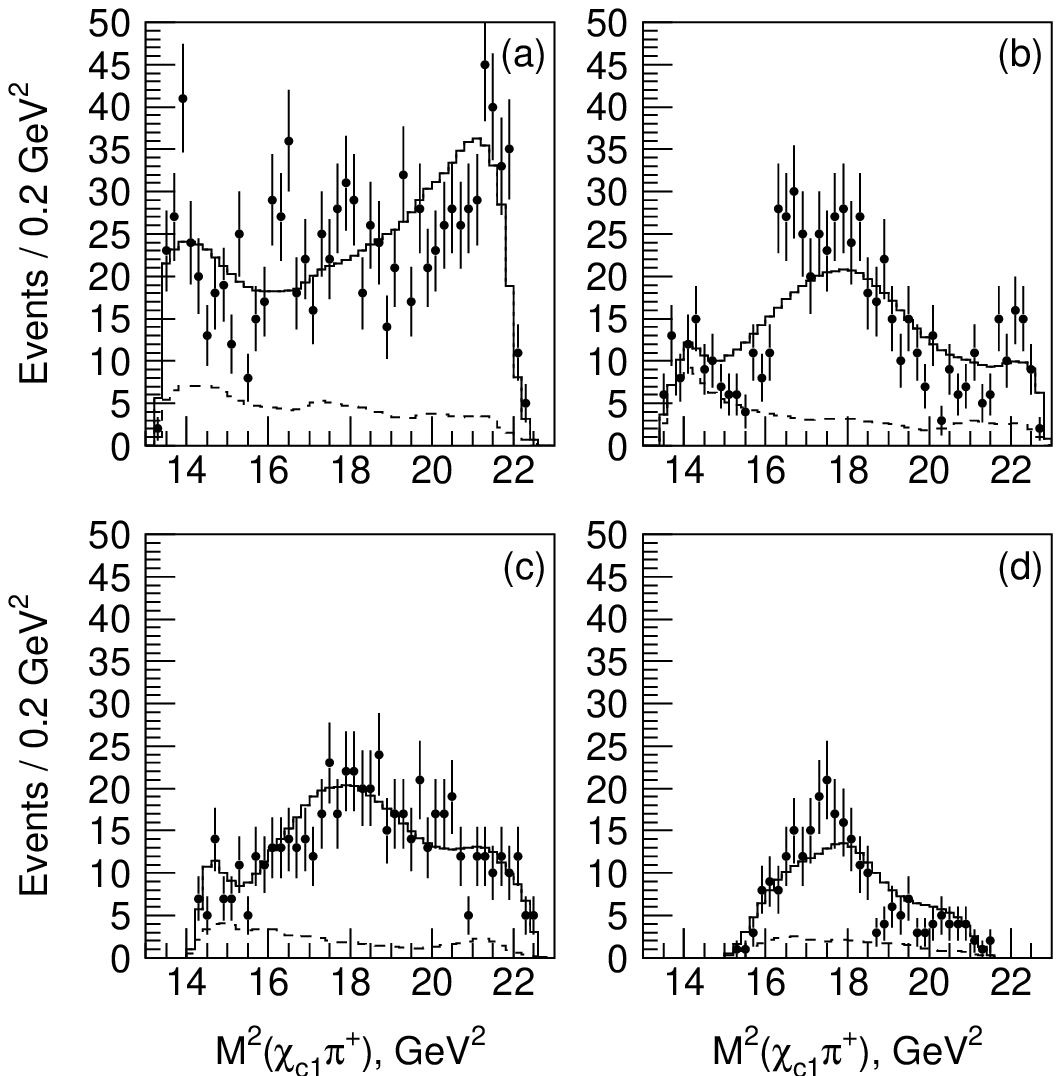}
\includegraphics[width=8cm]{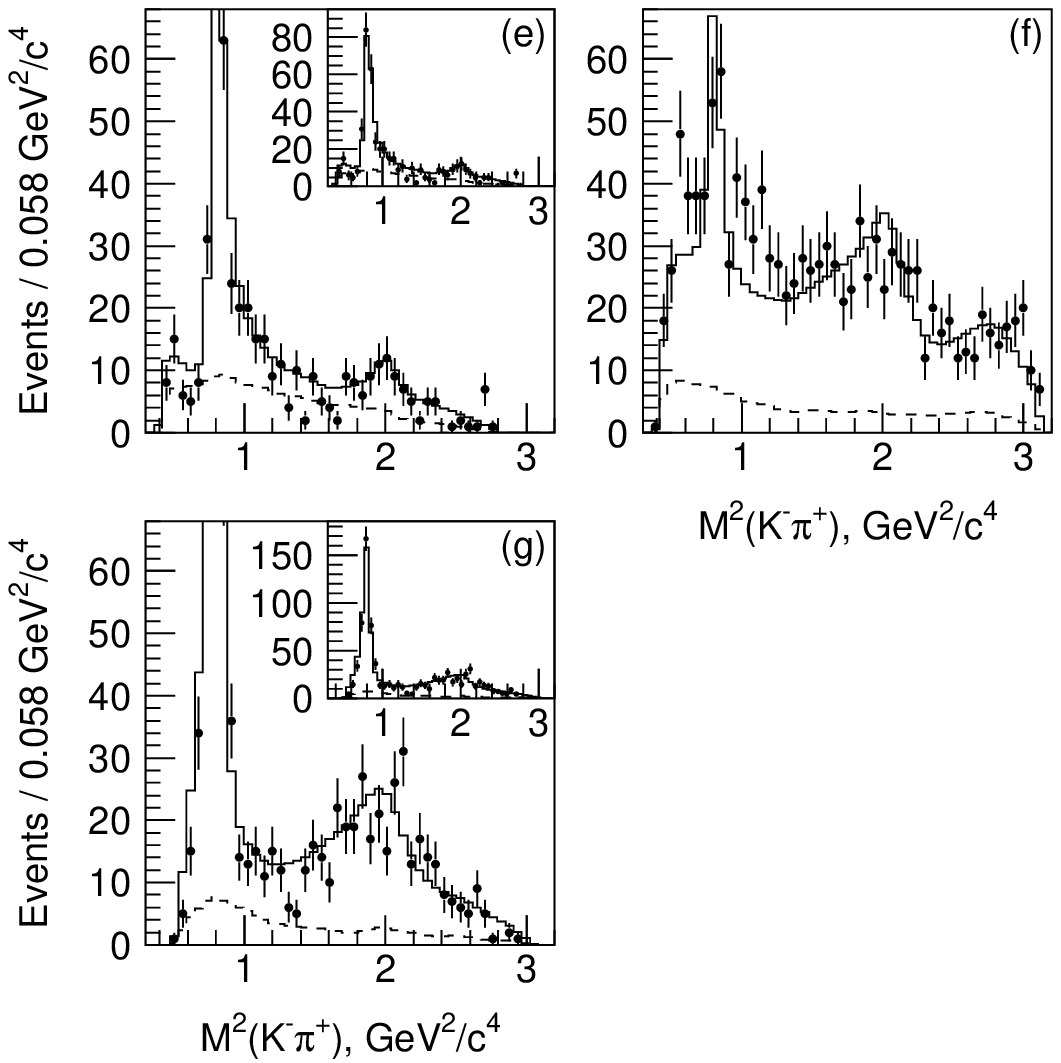}
\caption{ Dalitz plot fit without any $\z$ term. Projections are shown
  in (a)-(g) as described for Fig.~\ref{allk_s}. The dots with error
  bars represent data, the solid histograms are the fit results, and
  the dashed histograms represent the background. Here the fitting
  model with all known $\kst$ resonances and no resonance in the
  $\pip\ch$ channel is used. }
\label{allk_noz}
\end{figure}

The significance of the $\z$, calculated from the difference in log
likelihood between fits with and without the $\z$ contribution with
the change in the number of degrees of freedom taken into account, is
$10.7\,\sigma$.
The results of the fit with the $\z$ contribution excluded from the
default fit function are presented in Fig.~\ref{allk_noz}.
The fit fractions and significances for each of the resonances
included in the default model are listed in Table~\ref{fit_frac}.
\begin{table*}[htb]
\caption{The fit fractions and significances of all contributions for
  the fit models with a default set of $\km\pip$ resonances and one
  $\z$ or two $\z$ resonances.}
\label{fit_frac}
\renewcommand{\arraystretch}{1.2}
\begin{ruledtabular}
\begin{tabular}{c|rr|rr}
& \multicolumn{2}{c|}{One $\z$} & \multicolumn{2}{c}{Two $\z$} \\
\cline{2-5}
Contribution & Fit fraction & Significance & Fit fraction & Significance \\
\hline
$\z_{(1)}$     & $(33.1^{+8.7}_{-5.8})\%$ & $10.7\,\sigma$ & $( 8.0^{+3.8}_{-2.2})\%$ & $5.7\,\sigma$ \\
$\z_2$         & --                 & --            & $(10.4^{+6.1}_{-2.3})\%$ & $5.7\,\sigma$ \\
$\kappa$       & $( 1.9\pm1.8)\%$ & $ 2.1\,\sigma$ & $( 3.6\pm2.6)\%$ & $3.5\,\sigma$ \\
$\kst(892)$    & $(28.5\pm2.1)\%$ & $10.6\,\sigma$ & $(30.1\pm2.3)\%$ & $9.8\,\sigma$ \\
$\kst(1410)$   & $( 3.6\pm4.4)\%$ & $ 1.3\,\sigma$ & $( 4.4\pm4.3)\%$ & $2.0\,\sigma$ \\
$\kst_0(1430)$ & $(22.4\pm5.8)\%$ & $ 3.4\,\sigma$ & $(18.6\pm5.0)\%$ & $4.5\,\sigma$ \\
$\kst_2(1430)$ & $( 8.4\pm2.7)\%$ & $ 5.2\,\sigma$ & $( 6.1\pm2.9)\%$ & $5.4\,\sigma$ \\
$\kst(1680)$   & $( 5.2\pm3.7)\%$ & $ 2.2\,\sigma$ & $( 4.4\pm3.1)\%$ & $2.4\,\sigma$ \\
$\kst_3(1780)$ & $( 7.4\pm3.0)\%$ & $ 3.6\,\sigma$ & $( 7.2\pm2.9)\%$ & $3.8\,\sigma$ \\
\end{tabular}
\end{ruledtabular}
\end{table*}

To study the model dependence, we consider a variety of other fit
hypotheses. The results are summarized in Table~\ref{models}.
\begin{table*}[htb]
\caption{Different fit models that are used to study systematic
  uncertainties and the significances of the single- and double-$\z$
  hypotheses.}
\label{models}
\renewcommand{\arraystretch}{1.2}
\begin{ruledtabular}
\begin{tabular}{rlrrr}
& Model & Significance & One $\z$ vs. & Significance \\
&       & of one $\z$  & two $\z$     & of two $\z$ \\
\hline
 1 & default (see text)                        & $10.7\,\sigma$ & $5.7\,\sigma$ & $13.2\,\sigma$ \\
 2 & no $\kappa$                               & $15.6\,\sigma$ & $5.0\,\sigma$ & $16.6\,\sigma$ \\
 3 & no $\kst(1410)$                           & $13.4\,\sigma$ & $5.4\,\sigma$ & $14.8\,\sigma$ \\
 4 & no $\kst_0(1430)$                         & $10.4\,\sigma$ & $5.2\,\sigma$ & $14.4\,\sigma$ \\
 5 & no $\kst(1680)$                           & $13.3\,\sigma$ & $5.6\,\sigma$ & $14.8\,\sigma$ \\
 6 & no $\kst_3(1780)$                         & $12.9\,\sigma$ & $5.6\,\sigma$ & $14.4\,\sigma$ \\
 7 & add non-resonant $\ch\km$ term                     & $ 9.0\,\sigma$ & $5.3\,\sigma$ & $10.3\,\sigma$ \\
 8 & add non-resonant $\ch\km$ term, no $\kst(1410)$    & $11.3\,\sigma$ & $5.1\,\sigma$ & $13.5\,\sigma$ \\
 9 & add non-resonant $\ch\km$ term, no $\kst(1680)$    & $11.4\,\sigma$ & $5.3\,\sigma$ & $13.7\,\sigma$ \\
10 & add non-resonant $\ch\km$ term, no $\kst_3(1780)$  & $10.8\,\sigma$ & $5.4\,\sigma$ & $13.2\,\sigma$ \\
11 & add non-resonant $\ch\km$ term, release constraints on $\kappa$ mass \& width & $ 9.5\,\sigma$ & $5.3\,\sigma$ & $10.7\,\sigma$ \\
12 & add non-resonant $\ch\km$ term, new $\kst$ ($J=1$)   & $ 7.7\,\sigma$ & $5.4\,\sigma$ & $ 9.2\,\sigma$ \\
13 & add non-resonant $\ch\km$ term, new $\kst$ ($J=2$)   & $ 6.2\,\sigma$ & $5.6\,\sigma$ & $ 8.1\,\sigma$ \\
14 & LASS parameterization of S-wave component  & $12.4\,\sigma$ & $5.3\,\sigma$ & $13.8\,\sigma$ \\
\end{tabular}
\end{ruledtabular}
\end{table*}
The first row in Table~\ref{models} corresponds to the fit model with
the default set of $\km\pip$ resonances.
Rows 2 through 6 indicate the results from models in which one of the
$K\pi$ resonances: $\kappa$, $\kst(1410)$, $\kst_0(1430)$,
$\kst(1680)$ or $\kst_3(1780)$, respectively, is removed.
Row 7 shows the results when a non-resonant $\ch\km$ amplitude,
parameterized as $ae^{ib}e^{-cM(\ch\km)}$, where $a$, $b$ and $c$ are
free parameters, is added to the fit model.  This amplitude can be
related to a decay that proceeds via a virtual $B^*$.  Rows 8 through
10 show results from fits that include the non-resonant contribution,
but without the $\kst(1410)$, $\kst(1680)$ or $\kst_3(1780)$,
respectively.
Row 11 corresponds to a fit that includes the non-resonant term and
releases the experimental constraints on the mass and width of the
$\kappa$.
We also consider models that include the non-resonant contribution,
plus an additional $J=1$ (row 12) or $J=2$ (row 13) $\kst$ resonance
with floating mass and width.
Finally, we replaced the $\kappa$ and $\kst_0(1430)$ contributions
with the $S$-wave component parameterization suggested by the LASS
experiment~\cite{LASS} (row 14). 
We used the following form of the LASS
parameterization~\cite{LASS_BABAR}:
\begin{equation}
\begin{split}
A_0=F^{(1)}_B\cdot\frac{p_B}{m_B}\cdot
\Big(
&\frac{\sqrt{s}}{p\,(\cot\delta-i)}+\\
&e^{2i\delta}
\frac{m_0^2\,\Gamma_0/p_0}
{m_0^2-s-
i\,m_0\,\Gamma_0\frac{p}{p_0}\frac{m_0}{\sqrt{s}}}\Big).
\end{split}
\end{equation}
Here $s$ is the four-momentum-squared of the $\km\pip$ pair, $p$ is
the $\km$ momentum in the $\km\pip$ c.m.\ frame, $m_0$ is the mass and
$\Gamma_0$ is the width of the $\kst_0(1430)$, $p_0$ is the $\km$
momentum calculated for the pole mass of the $\kst_0(1430)$, and the
phase $\delta$ is determined from the equation
$\cot\delta=\frac{1}{ap}+\frac{bp}{2}$, where $a$, $b$ are the model
parameters. We used the LASS optimal values for the $a$ and
$b$~\cite{LASS_BABAR}.

\begin{figure}[htbp]
\includegraphics[width=8cm]{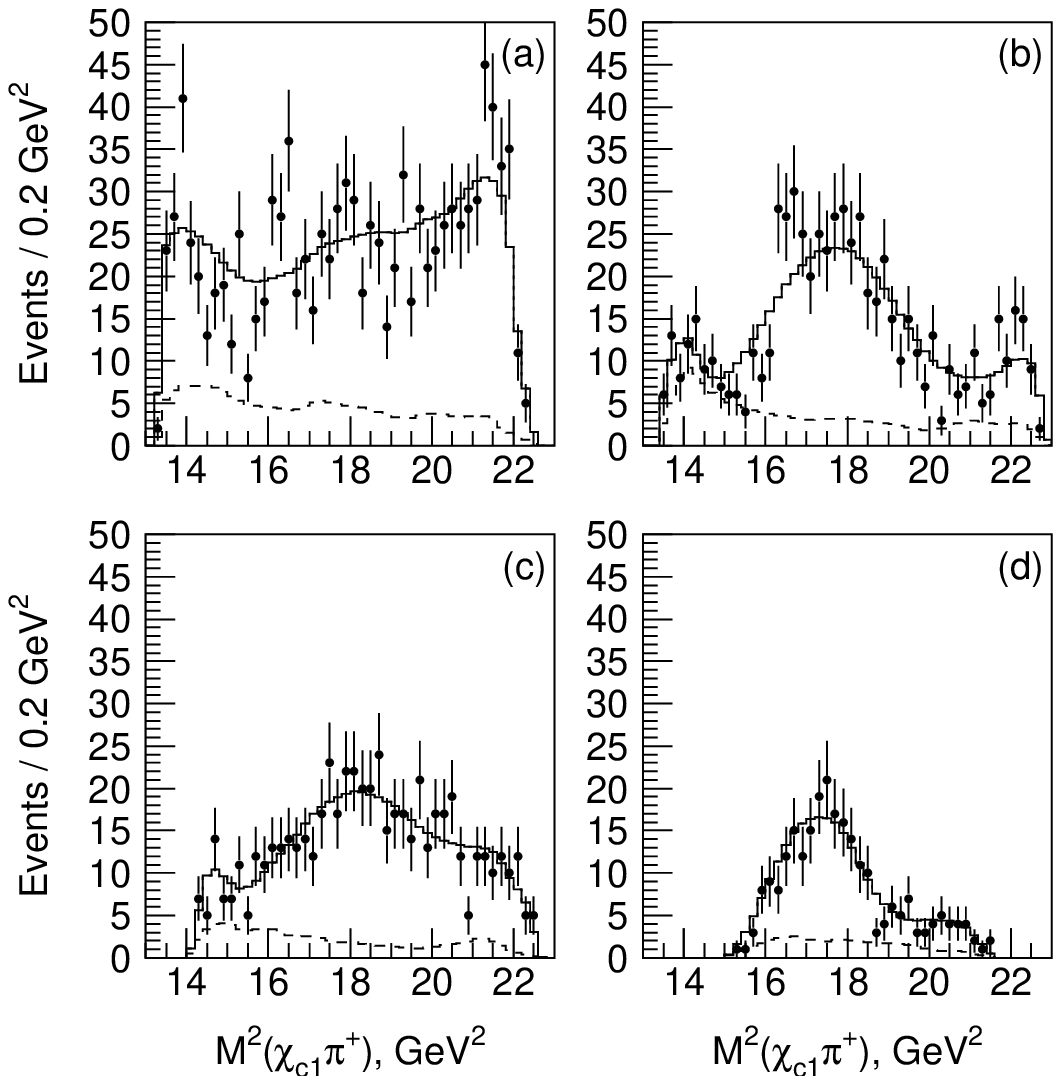}
\includegraphics[width=8cm]{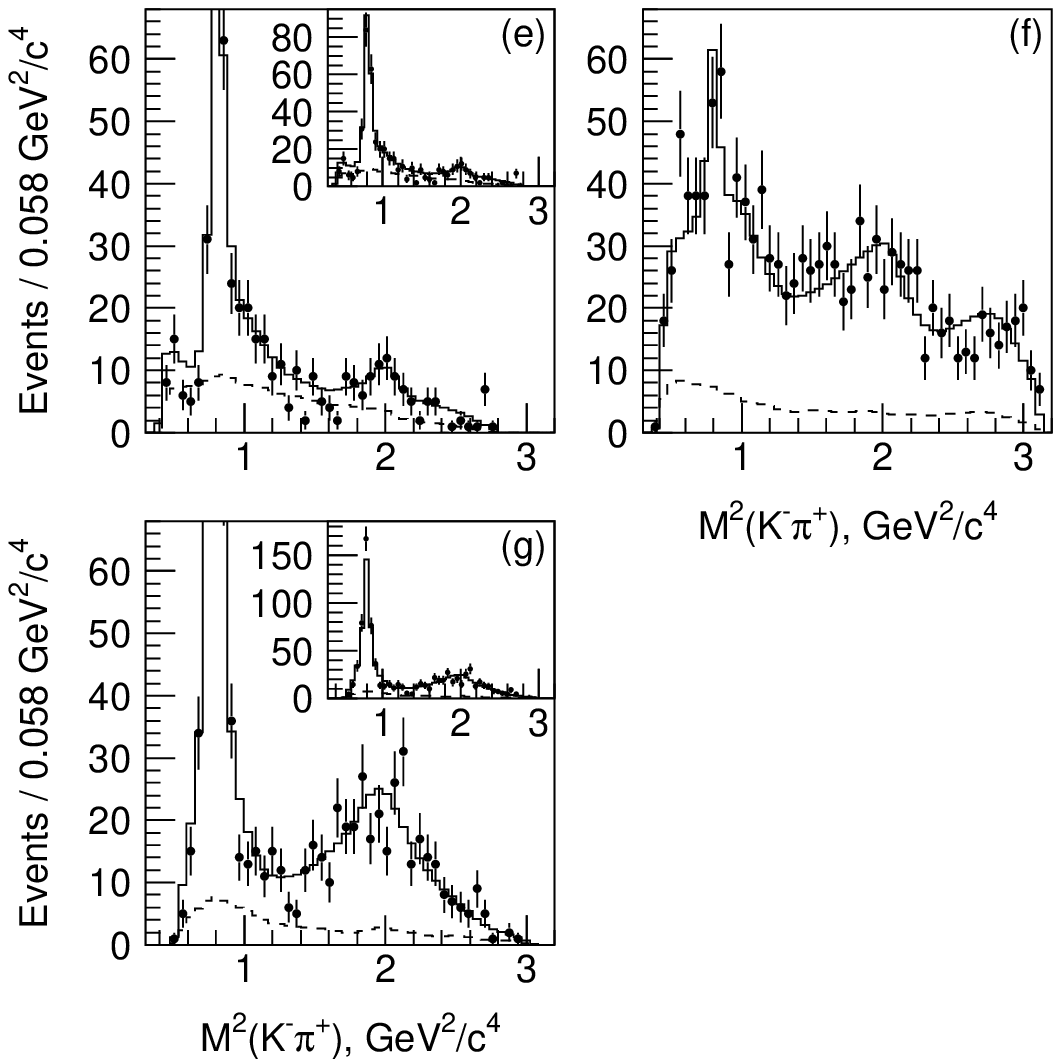}
\caption{ An alternative Dalitz plot fit without any $\z$ term.
  Projections are shown in (a)-(g) as described for Figs~\ref{allk_s}
  and~\ref{allk_noz}; point- and line-styles match those from
  Fig.~\ref{allk_noz}.  The fit model with all known $\kst$
  resonances, a $\ch\km$ non-resonant contribution, and a new
  $\kst_2$, but without a $\z$ term, is used. }
\label{allk_exp_j2_noz}
\end{figure}

For each fit model the $\z$ significance is estimated. The minimal
significance of $6.2\,\sigma$ corresponds to fit model 13 and is
considered as the $\z$ significance with systematics taken into
account. The fit result for model 13 without the contribution of the
$\z$ is shown in Fig.~\ref{allk_exp_j2_noz}.
For models with an additional $J=1$ or $J=2$ $\kst$ resonance with
floating mass and width, the resulting fitted masses and widths of the
additional $\kst$ resonances ($M=2.14\,\gevm$, $\Gamma=3.0\,\gev$ for
$J=1$ and $M=1.05\,\gevm$, $\Gamma =0.26\,\gev$ for $J=2$) do not
match those of any known $K\pi$ resonance~\cite{PDG}.

In the fits described above, the spin of the $\z$ is assumed to be
0. We find that the $J=1$ assumption does not significantly improve the
fit quality (in the default fit model, $-2\ln L$ changes from 17640.7
to 17638.3 for four additional degrees of freedom). 

It is not possible to distinguish the contributions of $\ch$ helicity
$+1$ and $-1$ in models where the spin of $\z$ is zero. Fits that
include both have nearly the same likelihood value as fits with only
one.

To address the question of fit quality we constructed a
two-dimensional histogram with varying bin sizes, in which there are a
minimum of 16 expected events in each bin (95 bins in total). A
$\chi^2$ is determined for this histogram, $\chi^2=\sum_i(n_i -
f_i)^2/f_i$, where $n_i$ is the number of events and $f_i$ is the
expectation value for the $i$-th bin, and a toy MC is used to
determine its confidence level. For the fit model with the default set
of the $\km\pip$ resonances and one $\z$ resonance (Fig.~\ref{allk_s})
the confidence level is 0.5\%. Such a low confidence level value
indicates that the shape of the structure is not well reproduced by a
single Breit-Wigner.
(The confidence levels of the fits without a $\z$ resonance, shown in
Figs~\ref{allk_noz} and \ref{allk_exp_j2_noz}, are $3\times10^{-10}$
and $9\times10^{-4}$, respectively.)

\section{Two $\z$'s?}

\begin{figure}[htbp]
\includegraphics[width=8cm]{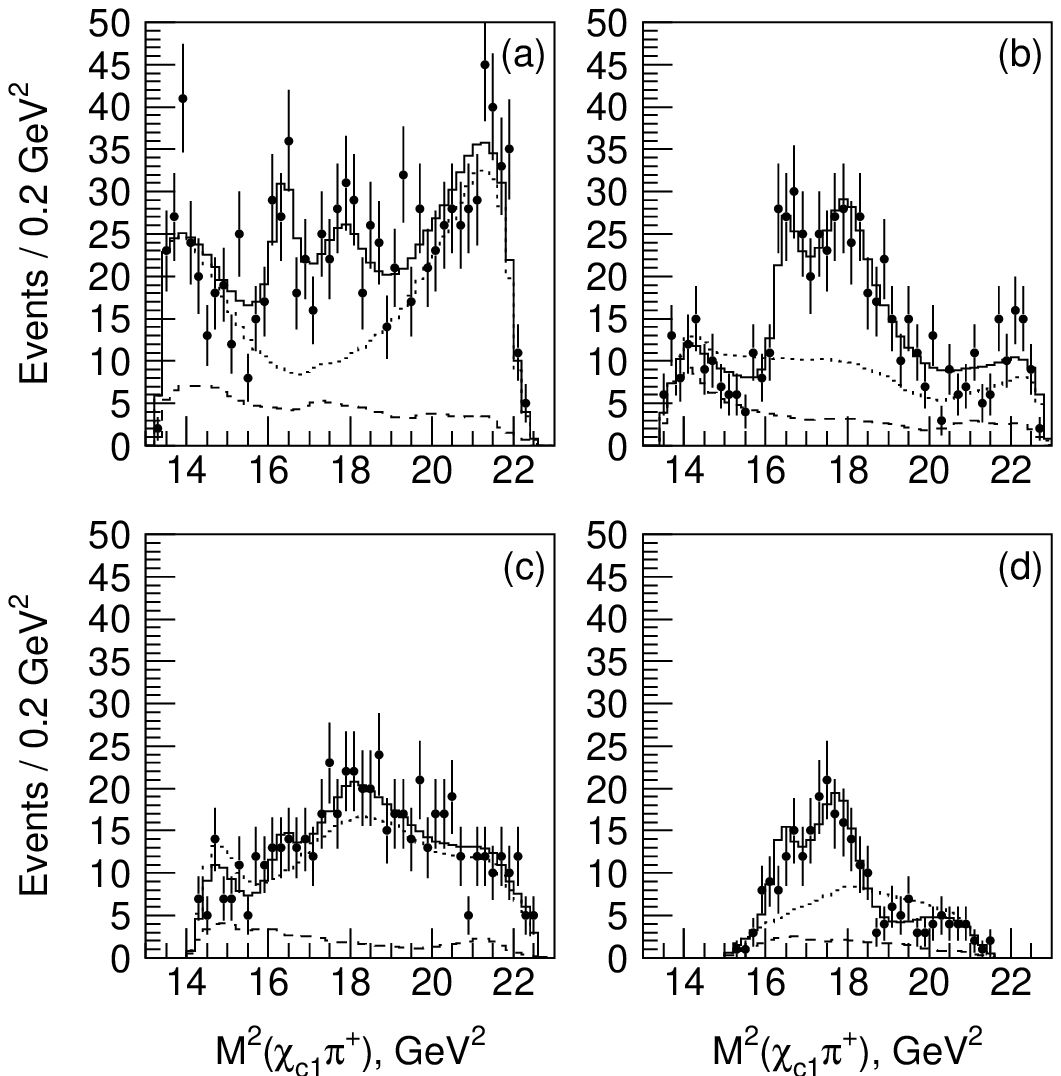}
\includegraphics[width=8cm]{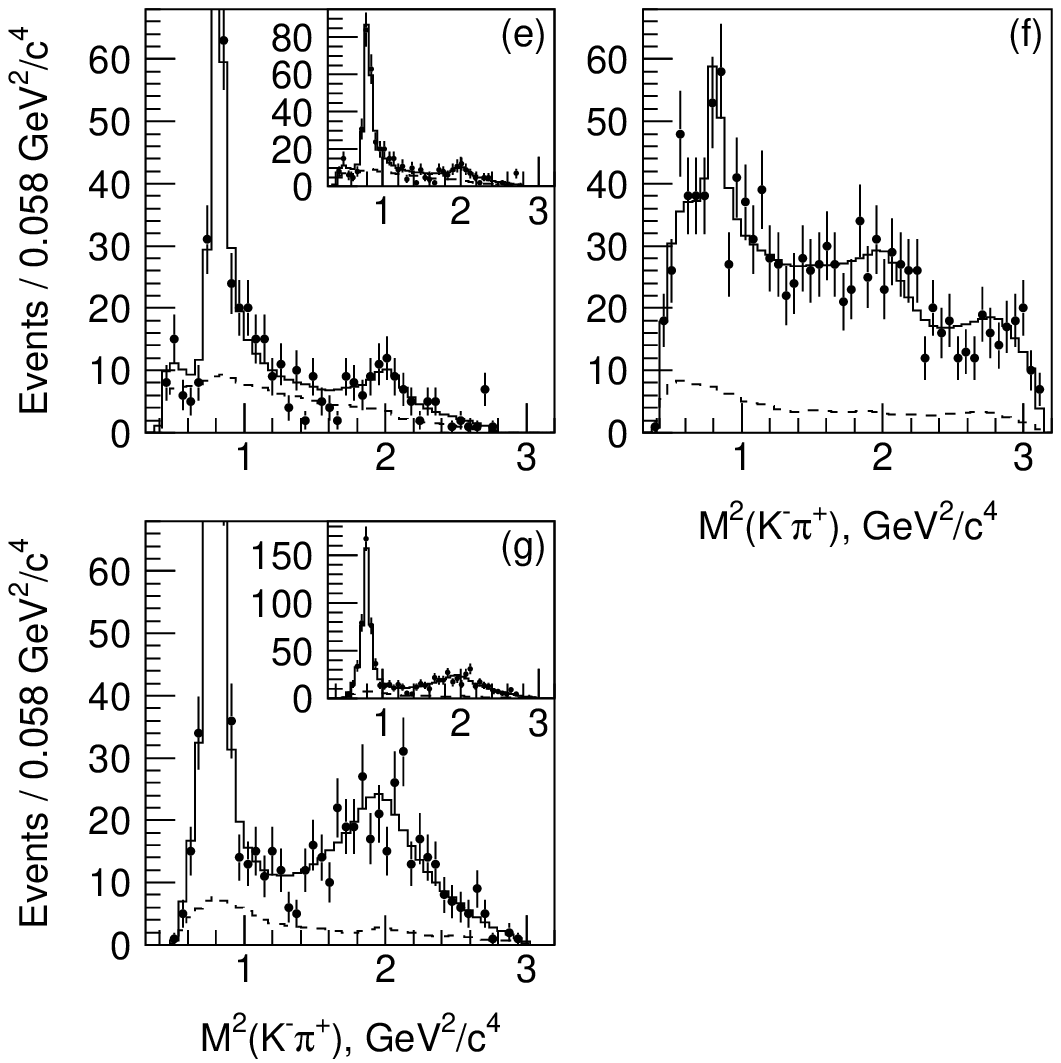}
\caption{ Dalitz plot fit including two $\z$ terms. Projections are
  shown in (a)-(g) as described for Fig.~\ref{allk_s}; point- and
  line-styles also match those of that figure. The fit model with all
  known $\kst$ resonances and two $\z$ terms ($\z_1$, $\z_2$) is
  used. }
\label{allk_s_s}
\end{figure}

In the Dalitz plot projections for the first and second vertical
slices ({\it cf.}\ the top two panels of Fig.~\ref{allk_s}) the
$M(\ch\pip)$ structure has a doubly peaked shape. This motivated us to
add a second $\z$ resonance to the default fit model.  The results of
the fit with this model are presented in Fig.~\ref{allk_s_s}.

The confidence level for this fit, calculated using the method used
for the one $\z$ hypothesis, is 42\%.
A comparison of the likelihood values for the one- and two-$\z$ fits
favors the two-$\z$ resonances hypothesis over the one-$\z$ resonance
scenario at the $5.7\,\sigma$ level.
The masses and widths of the two $\z$ resonances found from the 
two-$\z$ fit are
\begin{align*}
M_1 & =(\mone)\,\mevm,\\
\Gamma_1 & =(\gone)\,\mev,\\
M_2 & =(\mtwo)\,\mevm,\\
\Gamma_2 & =(\gtwo)\,\mev,
\end{align*}
with fit fractions of
$f_1=(8.0^{+3.8}_{-2.2}{^{+9.5}_{-4.2}})\%$ and
$f_2=(10.4^{+6.1}_{-2.3}{^{+51.5}_{-\phantom{1}0.7}})\%$.
The corresponding product branching fractions, calculated as
$\br(\B\to\km\pip\ch)\times f_{1,2}$, are
\begin{align*}
\br(\B\to\km\z_1)\times\br(\z_1\to\pip\ch)=\\
\bone,\\
\br(\B\to\km\z_2)\times\br(\z_2\to\pip\ch)=\\
\btwo.
\end{align*}
The product branching fractions are comparable to those of the
$Z(4430)^+$ and other charmonium-like states in a leading decay
mode~\cite{Z4430,PDG}.
The fit fractions and significances for each of the resonances
included in the model are listed in Table~\ref{fit_frac}.
We find that the phase difference between the two $\z$ resonances is
close to $\pi/2$: $\phi_{\z_2}-\phi_{\z_1}=1.7^{+0.2}_{-0.3}$.

To estimate systematic errors, we use the models listed in
Table~\ref{models}, with two $\z$ resonances instead of one, and
consider the maximum variations of the $\z _1$ and $\z _2$ masses,
widths and fit fractions for different fit models as a systematic
uncertainty. These uncertainties are given in the first row of
Table~\ref{zz_syst}.

The possibility of multiple minima can be an issue for complicated fit
models with many contributions. In light of this we randomly generated
the initial values for the fit parameters and repeated each fit
several times.  The deepest minimum is selected.
(This approach is used also for the single-$\z$ models.)
If any secondary minima are within $|\delta(2\ln L)|<2$ of the
selected solution, they are included in the systematic uncertainty
determination.

We also study the systematics due to the uncertainty in the form
factors for the decays. In addition to the default value of the $r$
parameter in the Blatt-Weisskopf parameterization $r=1.6\,\gev^{-1}$,
we also consider $r=1.0\,\gev^{-1}$ and $r=2.0\,\gev^{-1}$. The
variation of the $\z$ parameters is negligible. In addition, we change
the assumption about the value of the $\B$ decay orbital angular
momentum for those cases where several possibilities exist, as
discussed above. The resulting uncertainties are given in the second
row of Table~\ref{zz_syst}.

In the phase-space MC, the angular distributions of the
$\ch\to\jp\gamma$ and $\jp\to\leplep$ decays are assumed to be
uniform. To check the sensitivity of our results to this assumption,
we weight the MC events according to the expectations for the $\ch$
with zero helicity:
$1+2\cos^2\theta_{\ch}\cos^2\theta_{\jp}-\cos^2\theta_{\jp}$~\cite{helicity}. The
variation of the $\z$ parameters is found to be negligible.

We estimate systematic errors associated with the event selection by
repeating the analysis while loosening the selection requirements on
$M(\jp\gamma)$, $\mbc$ and track quality until the background level is
{\it increased} by a factor of two, and while tightening them until
the background level is {\it decreased} by a factor of two compared to
the nominal level. The systematic uncertainties estimated in this way
are given in the third row of Table~\ref{zz_syst}.

In the fits described above, the spins of both $\z$ resonances are
assumed to be zero.  We find that a $J=1$ assumption for either or both does
not significantly improve the fit quality, as shown in
Table~\ref{ll_j}, where we show results for all four possible
combinations of spin $J=0$ or $J=1$ assignment. The variations in the
$\z_1$ and $\z_2$ parameters for the different spin assignments are
considered as systematic uncertainties and are listed in the fourth
row in Table~\ref{zz_syst}.

To obtain the total systematic uncertainties, the values given in
Table~\ref{zz_syst} are added in quadrature.

\begin{table*}[htb]
\caption{Systematic uncertainties in the $\z$ mass, width and fit
  fraction due to fit model, uncertainty in the form factors,
  uncertainty in the $\ch$, $\jp$ decay angular distributions and
  variation of selection requirements.}
\label{zz_syst}
\renewcommand{\arraystretch}{1.5}
\begin{ruledtabular}
\begin{tabular}{l|ccc|ccc}
        & $M_1,\,\mevm$ & $\Gamma_1,\,\mev$ & $f_1,\,\%$ & $M_2,\,\mevm$ & $\Gamma_2,\,\mev$ & $f_2,\,\%$ \\
\hline
Fit model    & $^{+18}_{-18}$ & $^{+15}_{-\phantom{1}9}$  & $^{+4.6}_{-3.0}$ & $^{+27}_{-32}$ & $^{+97}_{-34}$ & $^{+18.5}_{-\phantom{1}0.7}$ \\
Formalism    & $^{+3}_{-0}$   & $^{+8}_{-0}$   & $^{+0.4}_{-\phantom{1.}0}$   & $^{+\phantom{1}0}_{-10}$  & $^{+\phantom{1}5}_{-11}$  & $^{+2.5}_{-\phantom{1.}0}$ \\
Selection    & $^{+\phantom{1}0}_{-21}$  & $^{+14}_{-\phantom{1}0}$  & $^{+2.4}_{-1.4}$   & $^{+87}_{-\phantom{1}0}$  & $^{+87}_{-\phantom{1}0}$ & $^{+6.1}_{-\phantom{1.}0}$ \\
Spin assignment & $^{+\phantom{1}7}_{-30}$  & $^{+42}_{-20}$ & $^{+8.0}_{-2.6}$ & $^{+156}_{-\phantom{1}10}$ & $^{+288}_{-\phantom{1}50}$  & $^{+47.6}_{-\phantom{11.}0}$ \\
\end{tabular}
\end{ruledtabular}
\end{table*}

\begin{table}[htb]
\caption{The $-2\ln L$ values and the change in the number of degrees
  of freedom for the fits with different spin assignments for the
  $\z_1$ and $\z_2$.}
\label{ll_j}
\renewcommand{\arraystretch}{1.2}
\begin{ruledtabular}
\begin{tabular}{cccc}
$J_1$ & $J_2$ & $-2\ln L$ & $\Delta d.o.f.$ \\
\hline
0 & 0 & 17599.2 & 0 \\
0 & 1 & 17594.3 & 4 \\
1 & 0 & 17597.5 & 4 \\
1 & 1 & 17590.1 & 8 \\
\end{tabular}
\end{ruledtabular}
\end{table}

In the extreme case, {\it i.e.} model 2 where the $\kappa$ is
eliminated, the two-resonance hypothesis is favored over the
one-resonance hypothesis with a $5.0\,\sigma$ significance. The
hypothesis with two $\z$ resonances is favored over the hypothesis
with no $\z$ resonances by at least an $8.1\,\sigma$ level for all
models.

We cross-check the estimated significances using toy MC. We generated
three types of toy MC events according to the fit results of the fit
model with the default set of $\km\pip$ resonances and with either
zero, one or two $\z$ resonances (100 samples of each type). We
perform the fits to these toy MC samples using the same three fit
models. The results (mean and r.m.s.) for the significance of the
single $\z$ resonance, the level at which the two-resonance hypothesis
is favored over the one-resonance hypothesis and the significance of
two resonances compared to the no-resonance hypothesis for all nine
combinations are given in Table~\ref{alan}.
\begin{table}[htb]
\caption{A comparison of the zero, one and two $\z$ resonance hypothesis
  for the toy MC samples generated for 0, 1 and 2 $\z$ resonances. The
  corresponding significances seen in the data are given for comparison.}
\label{alan}
\renewcommand{\arraystretch}{1.2}
\begin{ruledtabular}
\begin{tabular}{c|c|c|c|c}
Hypotheses & \multicolumn{3}{c|}{Toy MC samples} & \\
\cline{2-4}
compared & 0 & 1 & 2 & Data \\
\hline
1 over 0 & $(1.0\pm0.8)\,\sigma$ & $(9.1\pm1.0)\,\sigma$ & $ (9.4\pm0.9)\,\sigma$ & $10.7\,\sigma$ \\
2 over 1 & $(2.0\pm1.2)\,\sigma$ & $(1.3\pm0.8)\,\sigma$ & $ (5.4\pm1.0)\,\sigma$ & $5.7\,\sigma$ \\
2 over 0 & $(1.8\pm0.9)\,\sigma$ & $(8.8\pm1.0)\,\sigma$ & $(10.9\pm1.4)\,\sigma$ & $13.2\,\sigma$ \\
\end{tabular}
\end{ruledtabular}
\end{table}
We find that the pattern of the significances observed in data is
reproduced well by the toy MC with two $\z$ resonances.

\section{Branching fraction of the $\B\to\kst(892)^0\ch$ decay}

From the $\kst$ fit fraction from the two-$\z$ fit given in
Table~\ref{fit_frac}, we determine the branching fraction
\mbox{$\br(\B\to\kst(892)^0\ch)=\bkstch$}.
The systematic uncertainty is estimated in the same way as described
above for the $\z_{1,2}$ parameters. 
The result is significantly below the current world average
$(3.2\pm0.6)\times10^{-4}$~\cite{PDG}. However, this is the first
measurement of the branching fraction that takes into account
interference with other decay channels that produce the same final
state.
The fraction of longitudinal polarization is found to be
$(94.7^{+3.8}_{-4.8}{^{+4.6}_{-9.9}})\%$, which confirms the
conclusion that the $B\to\kst(892)\ch$ decay is dominated by
longitudinal polarization~\cite{soni,babar}.
The significances of other intermediate $\kst$ resonances are below
the $5\,\sigma$ level when systematic uncertainties from various fit
models are taken into account.

\section{Angular distributions of the $\ch$ and $\jp$ decays}

Angular distributions for $\ch\to\jp\gamma$ and $\jp\to\leplep$ decays
are not used in the Dalitz analysis and therefore provide a useful
cross-check.  For the $\ch$ in the helicity zero state the expected
angular distribution for $\ch\to\jp\gamma$ and $\jp\to\leplep$ decay
is
$P_0=\frac{9}{32}(1+2\cos^2\theta_{\ch}\cos^2\theta_{\jp}-\cos^2\theta_{\jp})$,
while for the $\ch$ in the helicity $\pm1$ state the expected angular
distribution is
$P_1=\frac{9}{32}(1-\cos^2\theta_{\ch}\cos^2\theta_{\jp})$.
Here it is assumed that different $\jp$ helicity states do not
interfere.
We integrate the helicity zero and helicity $\pm1$ components of the
fit function over the Dalitz plot and find the relative contributions
$w_0$ and $w_{\pm1}$. The expected angular distribution is then
$P=w_0P_0+w_{\pm1}P_{\pm1}$.

\begin{figure*}[htbp]
\includegraphics[width=14cm]{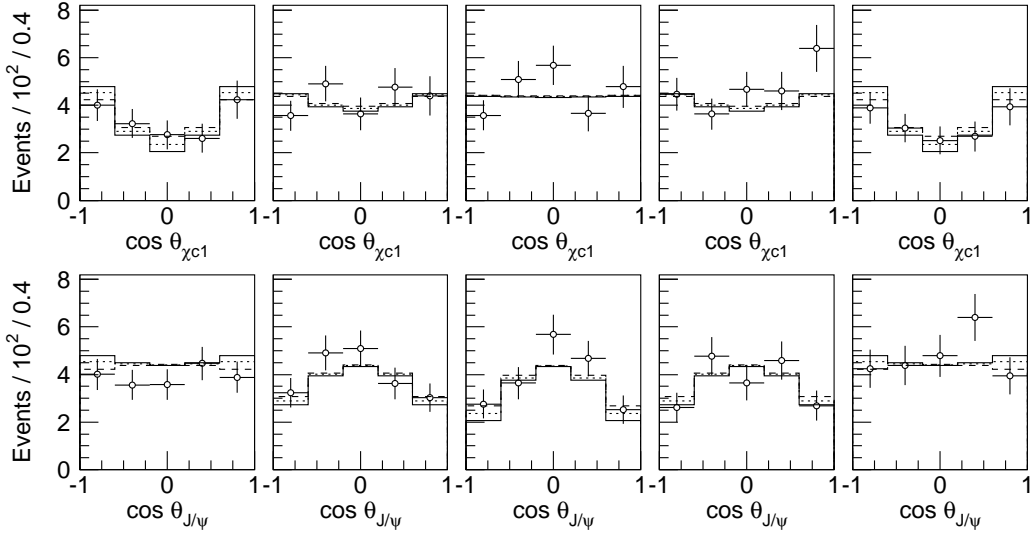}
\caption{ $\cos\theta_{\ch}$ ($\cos\theta_{\jp}$) distributions in
  $\cos\theta_{\jp}$ ($\cos\theta_{\ch}$) bins for the entire Dalitz
  plot. The dots with error bars are data, the solid (dashed)
  histograms are the predictions of the model with two $J=0$ ($J=1$)
  $\z$ resonances, the dotted histograms are the predictions of the
  model with no $\z$.  The $\cos\theta_{\jp}$ ($\cos\theta_{\ch}$)
  bins are ($-$1,$-$0.6), ($-$0.6,$-$0.2), ($-$0.2,0.2), (0.2,0.6) and
  (0.6,1).}
\label{all_super}
\end{figure*}

\begin{figure*}[htbp]
\includegraphics[width=14cm]{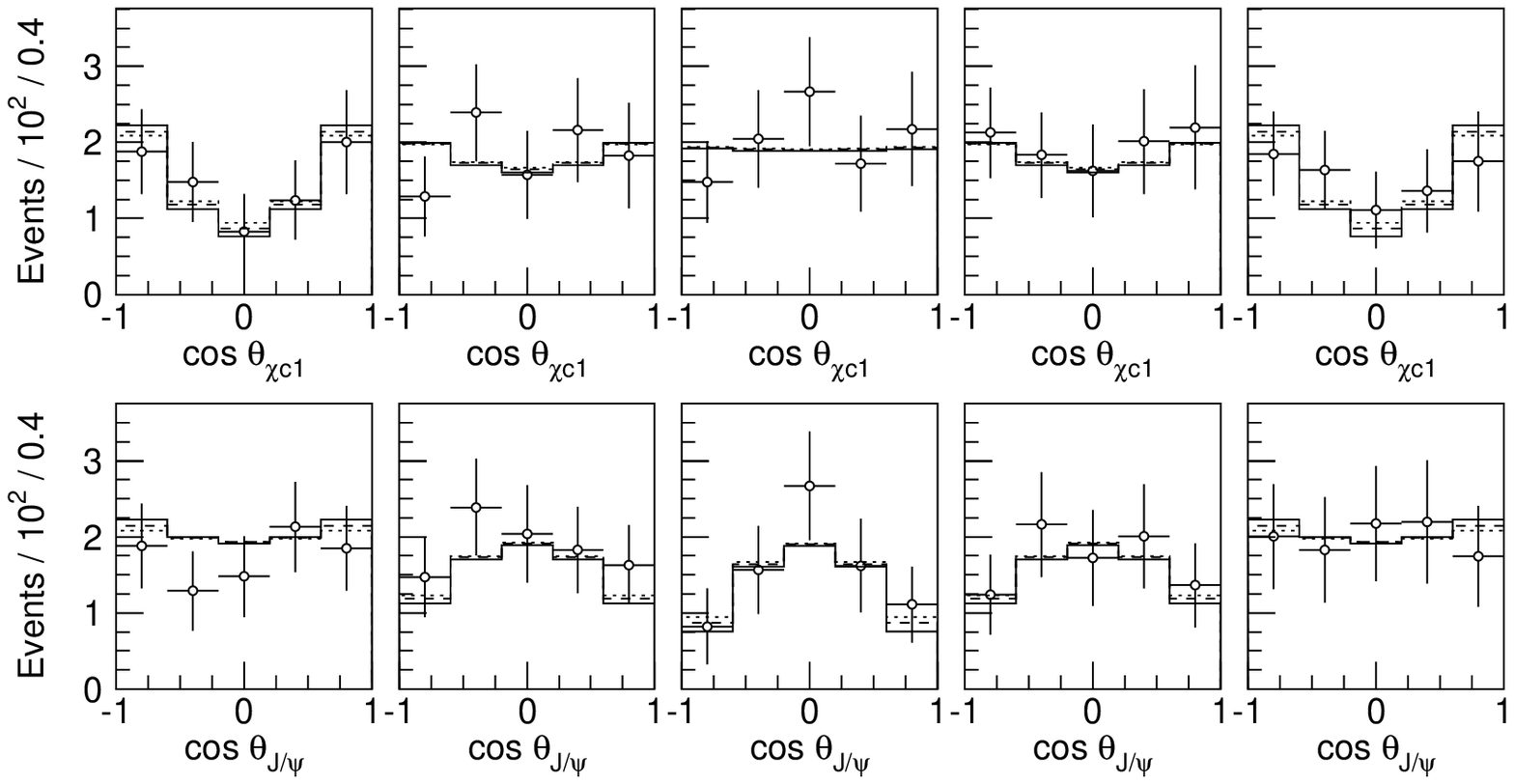}
\caption{ $\cos\theta_{\ch}$ ($\cos\theta_{\jp}$) distributions in
  $\cos\theta_{\jp}$ ($\cos\theta_{\ch}$) bins for the vertical Dalitz
  plot slice that contains the $\kst(892)$ signal. The dots with error
  bars are data, the solid (dashed) histograms are the predictions of
  the model with two $J=0$ ($J=1$) $\z$ resonances, the dotted
  histograms are the predictions of the model with no $\z$.  The
  $\cos\theta_{\jp}$ ($\cos\theta_{\ch}$) bins are ($-$1,$-$0.6),
  ($-$0.6,$-$0.2), ($-$0.2,0.2), (0.2,0.6) and (0.6,1).}
\label{1_5_9_super}
\end{figure*}

\begin{figure*}[htbp]
\includegraphics[width=14cm]{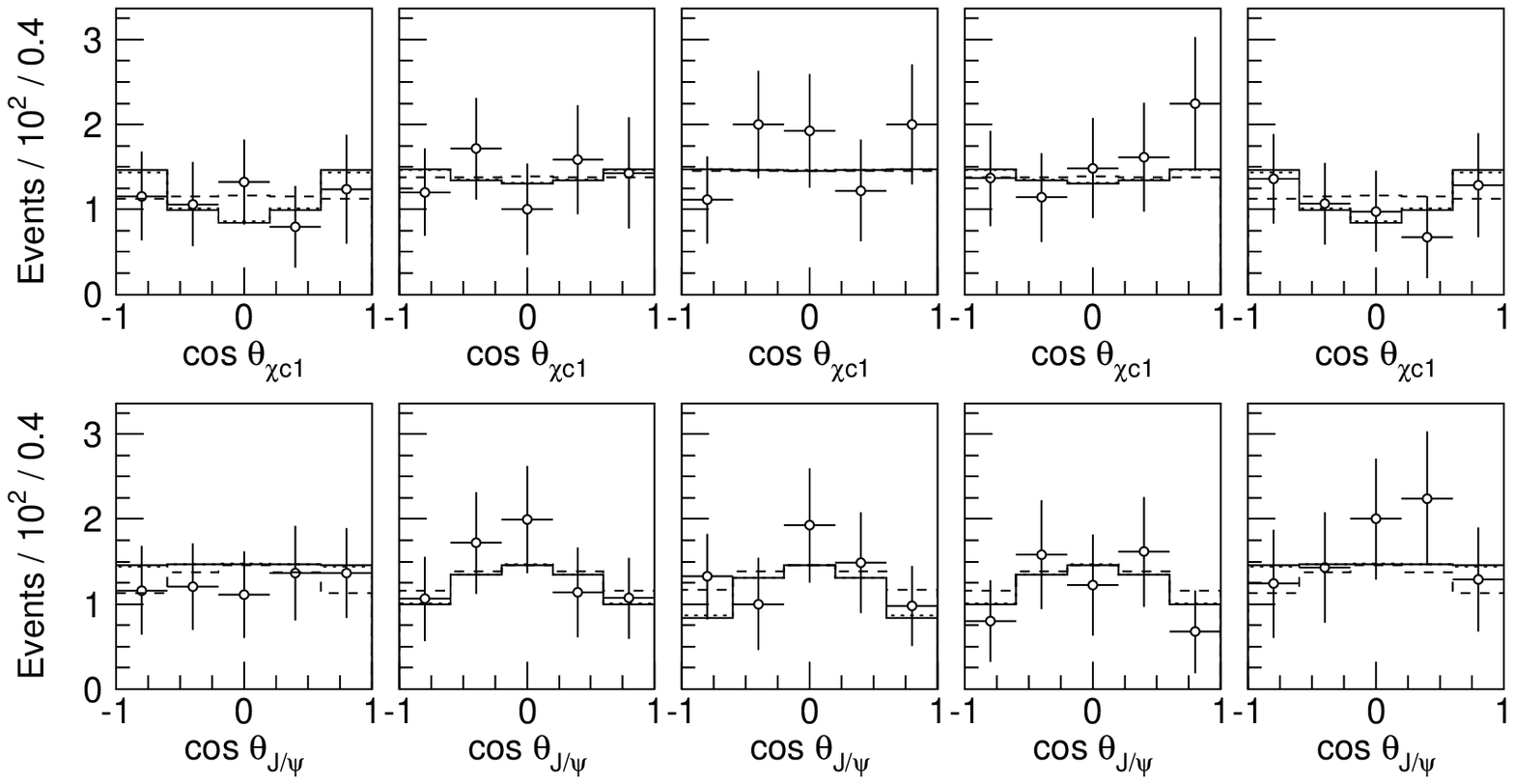}
\caption{ $\cos\theta_{\ch}$ ($\cos\theta_{\jp}$) distributions in
  $\cos\theta_{\jp}$ ($\cos\theta_{\ch}$) bins for the horizontal
  slice of the Dalitz plot that contains the $\z_{1,2}$ signals. The
  dots with error bars are data, the solid (dashed) histograms are the
  predictions of the model with two $J=0$ ($J=1$) $\z$ resonances, the
  dotted histograms are the predictions of the model with no $\z$.
  The $\cos\theta_{\jp}$ ($\cos\theta_{\ch}$) bins are ($-$1,$-$0.6),
  ($-$0.6,$-$0.2), ($-$0.2,0.2), (0.2,0.6) and (0.6,1).}
\label{6_7_8_super}
\end{figure*}

The $\cos\theta_{\ch}$ and $\cos\theta_{\jp}$ distributions for the
entire Dalitz plot, are presented in Fig.~\ref{all_super};
for the leftmost vertical slice containing the $\kst(892)$ signal in
Fig.~\ref{1_5_9_super};
and for the middle horizontal slice dominated by the $\z$ resonances
in Fig.~\ref{6_7_8_super}.
The agreement with predictions is good. It is evident that the different
models give very similar predictions and these angular
distributions are not useful for discriminating between them.

\section{Conclusions}

\begin{figure}[htbp]
\includegraphics[width=6cm]{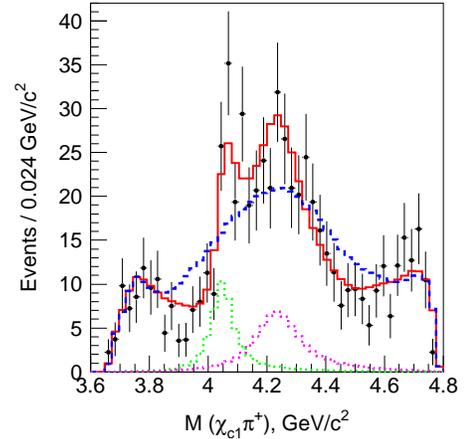}
\caption{ The $M(\ch\pip)$ distribution for the Dalitz plot slice
$1.0\,\gevms<M^2(\km\pip)<1.75\,\gevms$. The dots with error bars
represent data, the solid (dashed) histogram is the Dalitz plot fit
result for the fit model with all known $\kst$ and two (without any)
$\ch\pip$ resonances, the dotted histograms represent the contribution
of the two $\ch\pip$ resonances. }
\label{ped}
\end{figure}

A broad doubly peaked structure is observed in the $\pip\ch$ invariant
mass distribution in exclusive $\B\to\km\pip\ch$ decays.
When fitted with two Breit-Wigner resonance amplitudes, the resonance
parameters are
\begin{align*}
M_1 & =(\mone)\,\mevm,\\
\Gamma_1 & =(\gone)\,\mev,\\
M_2 & =(\mtwo)\,\mevm,\\
\Gamma_2 & =(\gtwo)\,\mev,
\end{align*}
with the product branching fractions of
\begin{align*}
\br(\B\to\km\z_1)\times\br(\z_1\to\pip\ch)=\\
\bone,\\
\br(\B\to\km\z_2)\times\br(\z_2\to\pip\ch)=\\
\btwo.
\end{align*}
The invariant mass distribution $M(\ch\pip)$ for the Dalitz plot slice
$1.0\,\gevms<M^2(\km\pip)<1.75\,\gevms$, where the contribution of the
structure in the $\pip\ch$ channel is most clearly seen, is shown in
Fig.~\ref{ped}.

Recently Belle observed the first candidate for a charmonium-like
state with non-zero electric charge, the $Z(4430)^+$~\cite{Z4430}. The
two resonance-like structures reported here represent additional
candidate states of similar character.
The existence of new resonances decaying into $\chi_{cJ}\pi$ is
expected within the framework of the hadro-charmonium
model~\cite{voloshyn}.

In addition, we measure the branching fractions
\mbox{$\br(\B\to\km\pip\ch)=\bkmpipch$} and
\mbox{$\br(\B\to\kst(892)^0\ch)=\bkstch$}.

%***** Acknowledgments *****
\begin{center}
{\bf Acknowledgments}
\end{center}
%----------- Long version, for most papers -----------
We thank the KEKB group for the excellent operation of the
accelerator, the KEK cryogenics group for the efficient
operation of the solenoid, and the KEK computer group and
the National Institute of Informatics for valuable computing
and SINET3 network support. We acknowledge support from
the Ministry of Education, Culture, Sports, Science, and
Technology of Japan and the Japan Society for the Promotion
of Science; the Australian Research Council and the
Australian Department of Education, Science and Training;
the National Natural Science Foundation of China under
contract No.~10575109 and 10775142; the Department of
Science and Technology of India;
the BK21 program of the Ministry of Education of Korea,
the CHEP SRC program and Basic Research program
(grant No.~R01-2005-000-10089-0) of the Korea Science and
Engineering Foundation, and the Pure Basic Research Group
program of the Korea Research Foundation;
the Polish State Committee for Scientific Research;
%-> remove for now: under contract No.~2P03B 01324;
the Ministry of Education and Science of the Russian
Federation and the Russian Federal Agency for Atomic Energy;
the Slovenian Research Agency;  the Swiss
National Science Foundation; the National Science Council
and the Ministry of Education of Taiwan; and the U.S.\
Department of Energy.


\begin{thebibliography}{99}

\bibitem{Z4430} S.-K.~Choi {\it et al.} (Belle Collaboration),
  Phys.\ Rev.\ Lett.\ {\bf 100}, 142001 (2008).

\bibitem{kekb} S.~Kurokawa and E.~Kikutani, Nucl.\ Instrum.\ Methods
  Phys.\ Res., Sect.\ A {\bf 479}, 117 (2002), and other papers
  included in this volume.

\bibitem{BELLE_DETECTOR} A.~Abashian {\it et al.} (Belle
  Collaboration), Nucl.\ Instrum.\ Methods Phys.\ Res., Sect.\ A {\bf
    479}, 117 (2002).

\bibitem{svd2} Z.~Natkaniec {\it et al.} (Belle SVD2 Group),
  Nucl.\ Instrum.\ Methods Phys.\ Res.\ Sect.\ A {\bf 560}, 1 (2006).

\bibitem{geant} R.~Brun {\it et al.}, GEANT 3.21, CERN DD/EE/84-1,
  1984.

\bibitem{PDG} W.-M.~Yao {\it et al.} (Particle Data Group),
  J.\ Phys.\ G {\bf 33}, 1 (2006).

\bibitem{cleo} S.~Kopp {\it et al.} (CLEO Collaboration),
  Phys.\ Rev.\  D {\bf 63}, 092001 (2001).

\bibitem{blatt-weisskopf} J. Blatt and V. Weisskopf, Theoretical
  Nuclear Physics, p.361, New York: John Wiley \& Sons (1952).

\bibitem{LASS} D.~Aston {\it et al.} (LASS Collaboration),
  Nucl.\ Phys.\  B {\bf 296}, 493 (1988).

\bibitem{LASS_BABAR}
  B.~Aubert {\it et al.} (BABAR Collaboration),
  %``Dalitz-plot analysis of the decays $B^\pm \to K^\pm \pi^\mp \pi^\pm$,''
  Phys.\ Rev.\  D {\bf 72}, 072003 (2005)
  [Erratum-ibid.\  D {\bf 74}, 099903 (2006)].

\bibitem{helicity} The $\ch\to\jp\gamma$ decay is governed by two
  helicity amplitudes, $H^{\ch}_{1,1}$ and $H^{\ch}_{0,1}$, where the
  first and second subscripts correspond to the helicity of the $\jp$
  and $\gamma$, respectively. Both experimental results and
  theoretical calculations indicate that the quadrupole contribution
  to the $\ch\to\jp\gamma$ transition is small, and therefore
  $|H^{\ch}_{1,1}|\simeq|H^{\ch}_{0,1}|$. \\
%
  C.~Baglin {\it et al.} (R704 Collaboration),
  %``ANGULAR DISTRIBUTIONS IN THE REACTIONS p anti-p $\to$ chi (1,2)
  %$\to$ gamma psi $\to$ gamma e+ e-,''
  Phys.\ Lett.\  B {\bf 195}, 85 (1987);\\
  K.~J.~Sebastian, H.~Grotch and F.~L.~Ridener,
  %``Multipole amplitudes in parity changing one photon transitions of
  %charmonium,''
  Phys.\ Rev.\  D {\bf 45}, 3163 (1992).

\bibitem{soni} N.~Soni {\it et al.}  (Belle Collaboration), Phys.\
  Lett.\ B {\bf 634}, 155 (2006).

\bibitem{babar} B.~Aubert {\it et al.}  (BABAR Collaboration),
  Phys.\ Rev.\  D {\bf 76}, 031102 (2007).

\bibitem{voloshyn} S.~Dubynskiy and M.~B.~Voloshin, arXiv:0803.2224 [hep-ph].

\end{thebibliography}
\end{document}